\documentclass[floatfix,aps,twocolumn,prx,superscriptaddress]{revtex4-1}
\usepackage{graphicx}
\usepackage{dcolumn}
\usepackage{bm}
\usepackage[T1]{fontenc}
\usepackage[latin9]{inputenc}
\usepackage{textcomp}
\usepackage{amsmath}
\usepackage{mathrsfs}
\usepackage{graphicx}
\usepackage{amssymb}
\usepackage{units}
\usepackage{xcolor}
\usepackage{float}

\newcommand{\iac}[1]{#1}
\newcommand{\ivan}[1]{#1}
\newcommand{\iaciac}[1]{#1}

\begin{document}

\title{Theory of the coherence of topological lasers}

\author{Ivan Amelio}
\affiliation{INO-CNR BEC Center and Dipartimento di Fisica, Universit{\`a} di Trento, 38123 Povo, Italy}
\author{Iacopo Carusotto}
\affiliation{INO-CNR BEC Center and Dipartimento di Fisica, Universit{\`a} di Trento, 38123 Povo, Italy}


\begin{abstract}
We present a theoretical study of the temporal and spatial coherence properties of a topological laser device built by including saturable gain on the edge sites of a Harper--Hofstadter lattice for photons.
\ivan{For small enough lattices the Bogoliubov analysis applies, the emission is nearly a single-mode one and the coherence time is  almost determined by the total number of photons in the device \iaciac{in agereement with the standard Schawlow-Townes phase diffusion. In larger lattices, looking at the lasing edge mode in}} the comoving frame of its chiral motion, the \iac{spatio-temporal} correlations of long-wavelength fluctuations display a Kardar-Parisi-Zhang (KPZ) scaling. \iac{Still, at very long} times, \iac{when} the finite size of the device starts to matter, the functional form of the \iac{temporal decay of coherence changes from the} KPZ stretched exponential to a \iac{Schawlow-Townes-like} exponential\iac{, while the nonlinear many-mode} dynamics of KPZ fluctuations \iac{remains visible as} an enhanced linewidth as compared to \iaciac{the single-mode Schawlow-Townes prediction.}
While we have established the above behaviors also for \iaciac{non-topological laser} arrays,
the crucial role of topology in protecting the coherence from static disorder is finally highlighted\iac{: our ground-breaking numerical calculations suggest the dramatically reinforced} coherence properties of topological lasers \iac{compared to} corresponding \iaciac{non-topological} devices. \iac{These results} open exciting possibilities for \iac{both} fundamental studies of non-equilibrium statistical mechanics and concrete applications to laser devices. 
\end{abstract}

\date{\today}

\maketitle

\section{{Introduction}}

The laser is one of the most fundamental tools in modern science \citep{siegman1987, svelto2010}. Its defining feature is the emission of radiation \iac{with} unprecedented long coherence length and time.
This makes laser sources essential ingredients in a wide range of applications \iac{and} justifies \iac{the} continuous theoretical and technological research of new devices. Also from a fundamental science perspectives, the  \iac{physical mechanisms underlying laser oscillation represent an} archetypical  model \iac{at the crossroad} of nonlinear physics, non-equilibrium statistical mechanics, and quantum optics~\cite{haken1983,gardiner2004,chiocchetta2017,keeling2017}.

Recent years have witnessed an explosion of the field of topological photonics~\citep{ozawa2019}.
Following \iac{the ground-breaking works}~\cite{haldane2008, wang2009}, experiments have initially \iac{focussed on demonstrating the topologically protected chiral propagation of light} along the edges of passive photonic lattices displaying topologically non-trivial band structures \iac{in a variety of different material platforms and frequency regions. In addition to the study of nonlinear topological photonics effects~\cite{smirnova2019}, a major attention has been devoted in the last years to laser operation in \iaciac{topological} edge modes~\cite{ota2020},} first in \iac{one-dimensional} SSH chains of polariton resonators \citep{stjean2017} and, \iac{soon afterwards,} in 2D photonic crystal \citep{bahari2017} and microcavity arrays \citep{bandres2018}. 

\ivan{Pioneering theoretical studies have focused on the  properties of the steady-state emission \iaciac{of topological laser devices}  \citep{harari2018,secli2019},    highlighting \iaciac{their} promise with respect to the power slope efficiency and to the robustness of the steady state emission} in the presence of a strong static defect \iac{located} on the edge \iac{of the lattice} or \iac{of a moderate disorder distributed throughout the device}.
However, a dynamical study of the temporal and spatial coherence properties \iac{of the resulting laser emission including the effect of noise due, e.g., to spontaneous emission} is still missing. \iac{The characterization of the ultimate limitations to the coherence properties is a key element in view of technological applications of topological lasers and will be the main topic of this work.}

In the absence of external noise sources, the long time coherence of a single-mode laser is ruled by the diffusion of the phase of the macroscopic electromagnetic field.
As it was first theoretically understood by Schawlow and Townes \citep{schawlow1958}, \iac{an intrinsic mechanism for} the diffusion of the phase \iac{is given} by spontaneous emission events\iac{, which result in an exponential decay of the temporal coherence function and in a corresponding broadening of the emission linewidth.}
\iac{In analogy with this ultimate linewidth of single-mode lasers, it is natural to investigate the ultimate limitations to the coherence of} a \iac{spatially extended} topological laser device \iac{whose physics involves a complex multi-mode nonlinear dynamics,}
\ivan{ to be treated with non-equilibrium statistical mechanics methods}. 

\iac{To facilitate the identification of the fundamental process common to all concrete realizations of topological laser device, we restrict to} the simplest scenario of a so-called class-A laser~\cite{longhi2018}, \iac{in which} the \iac{carrier dynamics in the gain medium can be neglected and laser operation is described in terms of} a coherent field amplified by \iac{a saturable gain medium with a temporally instantaneous response. With no loss of generality, we focus on the Harper--Hofstadter topological laser model and we consider that gain is distributed around the whole edge of the system~\citep{secli2019}. We further assume that the real part of the refractive index of the medium does not depend on light intensity. As usual in the semi-classical laser theory~\cite{gardiner2004,scully1997}, the effect of spontaneous emission is modeled as a spatio-temporally white noise in the stochastic differential equations for the multi-mode laser field.}

Since \iac{topological laser operation} occurs into a chiral mode localized on the one-dimensional edge of \iac{the two-dimensional} lattice, we expect that \iac{the coherence properties resemble the ones of one-dimensional} arrays of coupled laser resonators, for which it \iac{was anticipated in}~\citep{gladilin2014, altman2015, ji2015, he2015, lauter2017, squizzato2018}  that (for small intensity fluctuations) the long--wavelength phase dynamics follows a Kardar-Parisi-Zhang (KPZ) \iac{dynamics~\citep{kardar1986}
}  Here, we numerically prove that the same holds for the \iac{topological laser, once} the phase fluctuations are studied in the reference frame moving at the group velocity of the chiral mode. In particular, \iac{we show that} the spatio-temporal correlation functions \iac{satisfy the KPZ scaling} with no dramatic renormalization of the nonlinear coupling. 

\iac{While the KPZ scaling holds in spatially infinite systems or, in finite systems, until the dynamics has had time to experience the finite size, the study of key optical properties such as long-time coherence of a realistic topological laser device requires an explicit analysis of finite-size effects that goes beyond the available non-equilibrium statistical mechanics literature. In this respect, a main result of our study is that the very long time coherence of a finite system  does not follow the KPZ scaling but decays exponentially {\it {\`a} la} Schawlow-Townes, yet with a reinforced phase diffusion coefficient due to the intrinsic nonlinearity of the KPZ model.}
\ivan{For smaller lattices instead, the Bogoliubov analysis applies and yields a Petermann factor \cite{siegman1989a,siegman1989b} close to unity, meaning that the emission is nearly single-mode and, in agreement with the standard Schawlow-Townes picture, its coherence time is determined by the total number of photons in the device over the noise rate, which is the optimal scenario.}

\iac{\iaciac{Finally, as a dramatic consequence of the topological protection,} 
our ground-breaking numerical simulations suggest that coherence \iaciac{of a topological laser} is \iaciac{very weakly} affected by a \iaciac{static} disorder. \iaciac{Coherence is only lost when the strength of disorder} is on the order of the topological gap, that is orders of magnitude higher than what is needed to \iaciac{spoil} the coherence \iaciac{of} a non-topological array. This has a paramount importance in view of practical applications of topological lasers as sources of intense coherent light.}

The structure of the article is the following. In Sec.~\ref{sec:trivial} we focus on finite 1D laser chains and, after inspecting the crossover from \iac{the Kardar-Parisi-Zhang scaling} to \iac{a Schawlow-Townes-like phase diffusion} at very long times, we illustrate the significant reduction of the coherence time from the single-mode Schawlow--Townes prediction due to \iac{the nonlinear dynamics of the spatial fluctuations of the phase.}
In Sec.~\ref{sec:topo} we first review the basic concepts of \iac{topological laser operation into} the chiral edge mode of a 2D Harper--Hofstadter lattice. We then investigate \iac{the} emission spectrum and the one--dimensional character of \iac{the} phase fluctuations. In particular, we probe the universal KPZ \iac{scaling} of long wavelength fluctuations \iac{at intermediate times} and we compute the linewidth of the laser emission at very long times. In Sec.~\ref{sec:disorder} we demonstrate the robustness of \iac{the coherence of the topological laser emission} in the presence of \iac{a} static disorder. Conclusions and perspectives are finally drawn in Section \ref{sec:conclu}.

\section{One-dimensional array of laser resonators}
\label{sec:trivial}

\iac{In this Section we make use of a semi-classical approach based on stochastic differential equations to characterize the coherence properties of a simple model of spatially extended laser formed by a one-dimensional array of single-mode resonators. 
While a sizable literature has investigated the behaviour of long-wavelength fluctuations in terms of the KPZ equation~\citep{altman2015, gladilin2014,ji2015,he2015,lauter2017,squizzato2018}, not much attention has been paid to  the \iaciac{temporal dependence of the equal--space} correlation function of spatially finite systems, which is one of the experimentally most relevant quantities for lasers. To facilitate the reader, we will first set the stage by reviewing the Schawlow--Townes linewidth for single laser resonators and then we will move up to the case of interest of a finite array of coupled resonators.}

\subsection{Single resonator linewidth}

The basic features of laser operation \iac{in a single-mode resonator}, namely stimulated amplification, the \iac{spontaneous} breaking of the $U(1)$ symmetry of the field and \iac{the stabilization of the emission intensity by gain saturation}  are \iac{all captured} by the \iac{stochastic differential} equation
\begin{equation}
i \partial_t \psi =   \frac{i}{2} \left[ \frac{P}{1+ n/n_S}  - \gamma \right]  \psi + \sqrt{2D}\xi \ , 
\label{eq:single_resonator}
\end{equation}
\iac{for the single complex variable $\psi$ describing the amplitude of the electromagnetic field in the resonator,  $n=|\psi|^2$  is the field intensity, $\gamma$ is the loss rate of the ``cold'' resonator, $P$ the amplification rate induced by the (unsaturated) gain and $n_S$ is the gain saturation coefficient.}
The \iac{stochastic term} $\xi$ \iac{consists of a} white noise \iac{of unit variance} $\langle \xi^*(t) \xi(t') \rangle =  \delta(t-t')$ \iac{rescaled to have a} diffusion coefficient $D$.

Eq.~(\ref{eq:single_resonator}) \iac{is invariant under a $U(1)$ symmetry describing the rotation of the phase of the field $\psi$.} The steady--state field $\psi_0$ is zero  below the lasing threshold $P < P_{th} = \gamma$. \iac{Above the threshold $P > P_{th}$, it breaks the symmetry by choosing a specific phase and the steady-state intensity $n_0$} satisfies $n_0 = n_S \left[ \frac{P}{P_{th}}  - 1 \right]$. In particular, for $P =2 P_{th}$ it holds the very transparent expression $n_0 = n_S$.

To deal analytically with Eq.~(\ref{eq:single_resonator}) it is convenient to resort to the modulus--phase formalism: writing the field as  $\psi = \sqrt{n} e^{i\phi}$ one has
\begin{eqnarray}
\partial_t \phi &=&    \sqrt{\frac{D}{n}}\xi_1 \label{eq:phase_eq} \\
\partial_t n &=& \left[ \frac{P}{1+ n/n_S}  - \gamma \right]  n  + 2\sqrt{n D} \ \xi_2 \label{eq:density_eq}
\end{eqnarray}
\iac{in terms of the two real and independent} noises $\langle \xi_l(t) \xi_{l'}(t') \rangle = \delta_{l,l'} \delta(t-t'), \ l,l'=1,2$.
For small enough perturbations, the intensity relaxes to $n_0$ with \iac{at a} rate $\Gamma = \frac{\gamma(P-\gamma)}{P}$. \iac{On} the other hand, as a direct consequence of the $U(1)$ symmetry of the model,  the dynamics of the  phase is a diffusion with no restoring force.

Assuming noise is small enough to cause minor perturbations to the intensity, \iac{phase differences follow a Gaussian distribution, so within the so-called cumulant approximation \citep{gladilin2014}}, the autocorrelation of the field reads
\begin{equation}
g^{(1)}(t) = \langle \psi^*(t) \psi(0) \rangle \simeq n_0 e^{- \frac{1}{2} \langle [\phi(t) - \phi(0)]^2  \rangle}.
\label{eq:cumulant}
\end{equation}
\iac{As a consequence, the} usual Brownian motion scaling entails that the decay of coherence is described by the exponential 
\begin{equation}
g^{(1)}(t) \simeq n_0 e^{- \frac{\gamma_{ST}}{2}|t| }.
\label{eq:cumulant2}
\end{equation}
In Fourier space, this corresponds to a Lorentzian power spectral density. The coherence decay rate 
\begin{equation}
\gamma_{ST} = \frac{D}{n_0}.
\label{eq:Schawlow--Townes}
\end{equation}
 is the celebrated Schawlow--Townes linewidth \cite{schawlow1958,henry1982}. 
The crucial feature of this formula is the \iac{steady-state} intensity \iac{$n_0$} appearing at the denominator, \iac{which has the following physical interpretation}: the more photons \iac{are present} in the resonator, the less the phase of the field is perturbed when a \iac{additional} photon with a random phase is emitted into the \iac{mode by a spontaneous emission process}.

\subsection{Extended lasers and KPZ equation}
\label{ssec:trivial_g1}

\begin{figure*}[t]
\centering
\includegraphics[width=1\textwidth]{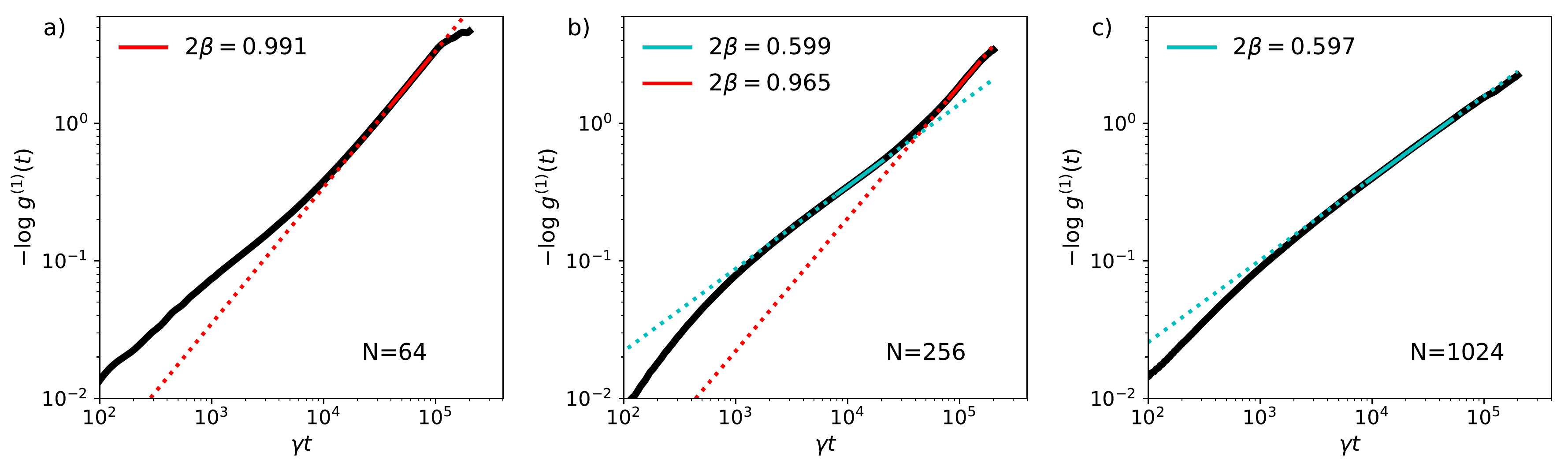}  
\caption{{\it \iac{Kardar-Parisi-Zhang to Schawlow-Townes crossover in the temporal coherence of finite, one-dimensional laser arrays}}  
\iac{The numerical prediction of Eq.~(\ref{eq:array_resonators}) for the} logarithm of the equal--space time correlation function $-\log g^{(1)}(t)$ is \iac{plotted} in loglog scale \iac{as a function of time for increasing system sizes $N_x=64$ (left), $N_x=256$ (middle), $N_x=1024$ (right). In all cases, lasing occurs around $k_x^{las}=0$. \iac{For a given} temporal window, the decay of the coherence is dominated by a Schawlow--Townes--like diffusion $g^{(1)}(t) \sim e^{-B|t|}$ for small sizes [panel (a)] and by the KPZ behavior $g^{(1)}(t) \sim e^{-B|t|^{2/3}}$ for large sizes [panel (c)]. The  crossover between the two regimes is visible for intermediate sizes [panel (b)]. The cyan and red lines are linear fits of $\log[-\log g^{(1)}(t)]$ with $2\beta \log|t| + B'$. Other system parameters: $J=0.5 \gamma$, $P=2\gamma$, $n_S=1000, D=\gamma$.}
}
\label{fig:crossover_KPZ_ST}
\end{figure*}

For a \iac{single single-mode} resonator, the linewidth is inversely proportional to the number of photons \iac{present in the mode}. For an array of coupled resonators, the naive intuition is that \iac{the linewidth will be determined by} the number of photons \iac{that are present} within some correlation length. In particular, for small system sizes all resonators oscillate in phase   and, for a given number of photons per resonator $n_0$, 
the coherence time is proportional to length of the array. \iac{On the other hand,} for large enough systems, it is known that the physics is described by the Kardar--Parisi--Zhang (KPZ) equation\citep{he2015}. \iac{The main goal of this Section is to connect these two points of view into a unified perspective.}

\iac{To attack this question in a quantitative way, we make use of a generalization of} Eq.~(\ref{eq:single_resonator}) to a one--dimensional  array of $N_x$ coupled laser resonators $x=1,...,N_x$
\begin{equation}
i \partial_t \psi_x = - J \psi_{x+1} - J \psi_{x-1} + \frac{i}{2} \left[ \frac{P}{1+ \frac{n_x}{n_S}}  - \gamma \right]  \psi_x + \sqrt{2D} \xi_x
 \ , 
\label{eq:array_resonators}
\end{equation}
with \iac{inter-site} coupling $J$ and \iac{independent noises} $\langle \xi_x^*(t) \xi_{x'}(t') \rangle =  \delta_{xx'} \delta(t-t')$. Periodic boundary conditions are assumed.

For \iac{large $J$} and calling $a$ the distance between neighbouring resonators, one can interpret  Eq.~(\ref{eq:array_resonators}) as a discrete version of  a continuous field of mass \iac{$m = \frac{1}{2Ja^2}$.}
The corresponding continuous equation is the complex Ginzburg--Landau equation \citep{wouters2007}. Assuming fast relaxation of the intensity fluctuations, \iac{one can focus on the dynamics of the phase (derivation reviewed in Sec. SI \cite{suppl}), which} is described by the Kuramoto--Sivashinsky  equation (KSE)
\begin{equation}
\partial_{{t}} {\phi} =  \frac{1}{2m} \left[ -\frac{\Gamma^{-1}}{2m} \partial^4_{{x}
} {\phi} 
- (\partial_{{x}}   {\phi} )^2 \right] + \sqrt{\frac{D}{n_0}} {\xi_1}.
\end{equation}  
\iac{Here,} $\phi$ is the unwound phase living on the real axis, and not the compact one restricted to $[0,2\pi]$. 

The characteristic scales of the system as a function of the microscopic parameters have been reported in   \cite{gladilin2014}:
$$ l^* = \left[ \frac{J^4 }{\Gamma^3 Dn_0^{-1}}  \right]^{1/7}, \ \  t^* = \left[ \frac{J^2}{\Gamma^5 (Dn_0^{-1})^4}  \right]^{1/7}, $$
\begin{equation}
 \ \  \phi^* = \left[ \frac{(Dn_0^{-1})^2}{J \Gamma}  \right]^{1/7}. \label{eq:star_units}
\end{equation}
Measuring space, time and (unwound) phase  in terms of $l^*, t^*, \phi^*$ the adimensional KSE reads
\begin{equation}
\partial_{\tilde{t}} \tilde{\phi} = - \partial^4_{\tilde{x}
} \tilde{\phi}  - (\partial_{\tilde{x}}   \tilde{\phi} )^2 + \tilde{\xi}.
\label{eq:KSE_tilde}
\end{equation}
\iac{However, since other scales can be relevant for} the $2D$ topological laser to be discussed below, in most of the paper we will use the physical (without tilde) space, time  and phase \iac{variables}. A rescaling will be proposed in order to study KPZ features in Sec.~\ref{ssec:kpz}.



The renormalization group analysis  shows that at long distances and times the KSE flows to the KPZ universality class\citep{ueno2005}. 
The KPZ equation \citep{kardar1986} reads 
\begin{equation}
\partial_{\tilde{t}} \tilde{\phi} =  \nu \partial^2_{\tilde{x}
} \tilde{\phi}  + \frac{\lambda}{2}(\partial_{\tilde{x}}   \tilde{\phi} )^2  + \sqrt{\mathcal{D}} {\xi_1}
\label{eq:KPZ}
\end{equation}
and it was originally proposed to describe the growth of interfaces.
Its scaling behavior at low energies (and assuming an infinite system and stationary regime) is characterized by  the correlation function
\begin{equation}
\Delta\tilde{\phi}_{\tilde{x},\tilde{t}}^2 \equiv \langle [\tilde{\phi}(\tilde{x},\tilde{t}) - \tilde{\phi}(0,0)]^2  \rangle
\end{equation}
and by two exponents $\chi, z$ which determine the asymptotic behavior of the spatial and temporal correlations respectively, according to $\Delta\tilde{\phi}_{\tilde{x},0}^2 \sim \tilde{x}^{2\chi}$ and $\Delta\tilde{\phi}_{0,\tilde{t}}^2 \sim \tilde{t}^{2\chi/z}$. 
In $1D$ we have $\chi=1/2$ for the roughness exponent  and $z=3/2$ for the dynamical \iac{exponent}; even more precisely it holds
\begin{equation}
\Delta\tilde{\phi}_{\tilde{x},\tilde{t}}^2=\left( \frac{1}{2}\lambda A^2 t \right)^{2/3} g_{KPZ}\left(\frac{\tilde{x}}{(2\lambda^2 A \tilde{t}^2)^{1/3}} \right)
\label{eq:scalingKPZ}
\end{equation}
where $A = \mathcal{D}/\nu$ is the variance of $\partial_{\tilde{x}}   \tilde{\phi}$.
The universal function $g_{KPZ}$ is known exactly \citep{prahofer2004} and we \iac{here recall its limiting values} $g_{KPZ}(u) - 2|u| \to 0$ for $ u \to \infty$ and $g_{KPZ}(u)\to 1.150...$ for $u \to  0$.

As a consequence, the equal--time correlation function has the random walk form $\Delta\tilde{\phi}_{\tilde{x},0}^2=\frac{A}{2}|\tilde{x}|$, for which the KPZ nonlinearity $\lambda$ \iac{does not} play any role. In other words, only looking at the spatial correlations the dynamics is not distinguishable from the \iac{one of the linear ($\lambda=0$) Edwards--Wilkinson (EW) model. In both cases, one has $\chi=1/2$ which corresponds to an exponential decay of the spatial coherence~\cite{gladilin2014,he2015}.}

\iac{A difference is instead visible in the spatio-temporal correlations, which have $z=2$ in the linear EW model} and $z=3/2$ in KPZ. In fact, at small distances and times, the nonlinear term in the KSE can be neglected and \iac{a linearized Bogoliubov analysis is accurate. \iac{A numerical illustration of the collapse of the coherence functions to the scaling form (\ref{eq:scalingKPZ}) will be given later on in Fig.\ref{fig:KPZ_scaling}(b)} in comparison with the topological lasing case.
Even though it goes beyond the present work, it is interesting to remind that the crossover from EW to KPZ physics was shown in~\cite{ji2015} to be slower in the presence of a nonlinear refractive index.}

\subsection{Linewidth of extended lasers}
\label{subsec:linewidth_trivial}

\begin{figure}[t]
\centering
\includegraphics[width=1\columnwidth]{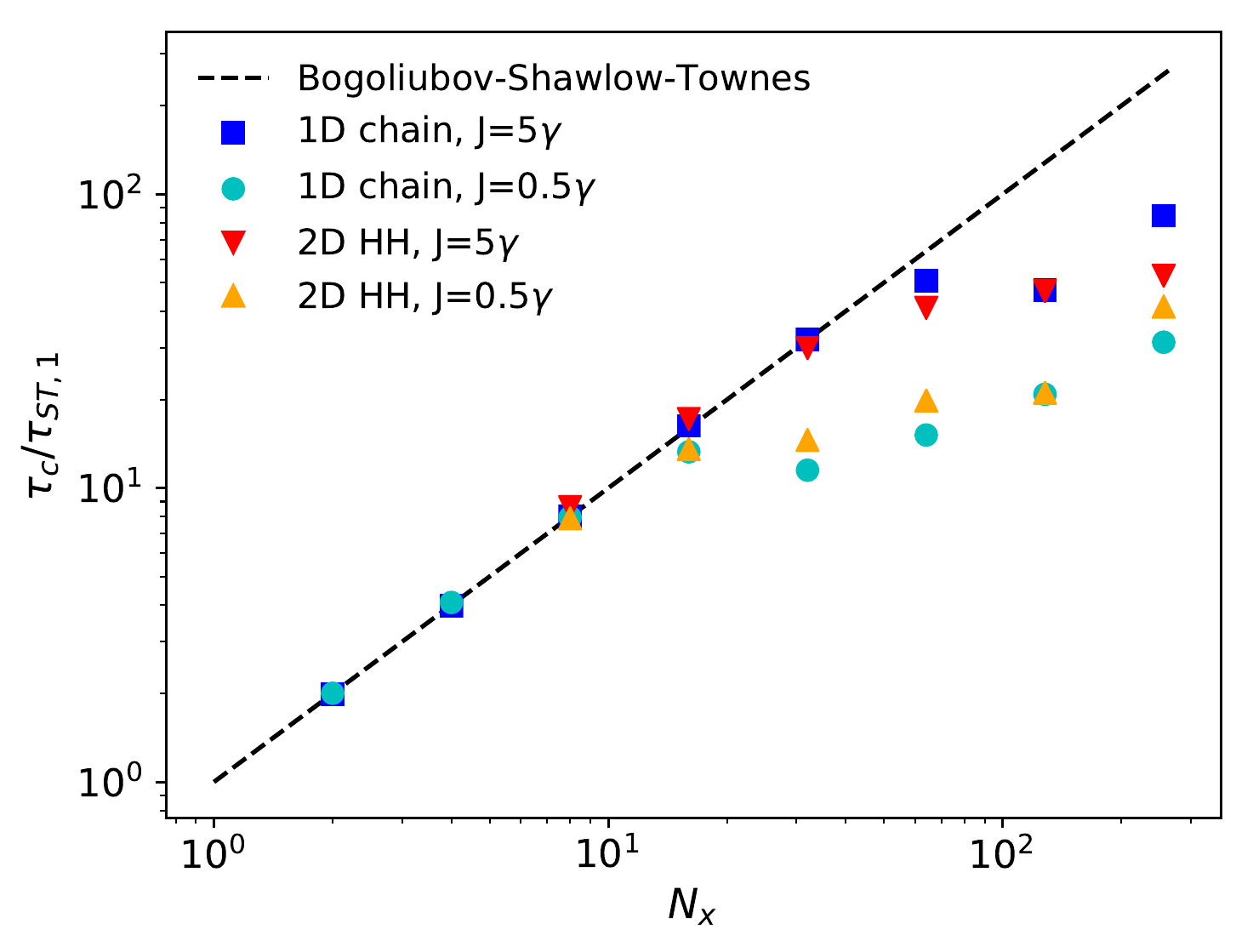}
\caption{ {\it Scaling of the coherence time with \iac{system size} $N_x$.}  
The coherence time $\tau_c$ is extracted \iac{from the long-time exponential decay of coherence for different systems. Blue and cyan markers refer to the one-dimensional, topologically trivial arrays of laser resonators of Eq.~(\ref{eq:array_resonators}) with different values of the inter-site coupling $J=5\gamma$ and $J=0.5\gamma$. The red and orange markers refer to the topological laser case of Eq.~(\ref{eq:topo_laser}) for the same two values of $J$. For each point, the value of the coherence time $\tau_c$ is normalized to the single-site coherence time $\tau_{ST,1} \equiv \tau_{ST}/N_x$ where $\tau_{ST}$ is the Bogoliubov--Schawlow-Townes prediction [(\ref{eq:Schawlow--Townes_xN}) for the trivial and (\ref{eq:BogoDrift}) for the topological device].}
}
\label{fig:Nscaling}
\end{figure}

\iac{The KPZ results reviewed in the previous sub-section apply to spatially infinite systems in the long-time limit, but are not of direct applicability to the concrete problem of the emission linewidth of a realistic laser device which necessarily has a finite spatial extension. To this purpose, the crucial quantity to study is the time dependence of the} equal--space correlator 
\begin{equation}
g^{(1)}(t) = \frac{1}{n_0} \left|  \langle \psi^*(x, t) \psi(x, 0) \rangle  \right|,
\end{equation}
\iac{which characterizes the temporal coherence of the emission. The dependence on $x$ has been dropped since we are considering a spatially uniform system.} 

\subsubsection{Linearized Bogoliubov prediction}

\iac{In the Bogoliubov limit, where the linearization on top of the noiseless state is a good approximation, modes of different momenta are decoupled. For a discrete lattice, one obtains with simple algebra that}
\begin{multline}
\langle [\phi(x,t) - \phi(x,0)]^2  \rangle =
\frac{1}{N_x} \sum_x \langle [\phi(x,t) - \phi(x,0)]^2  \rangle = \\
= \frac{1}{n_0N_x} \sum_k \bar{D}_k \int_0^t dt' e^{-2i\omega_k (t-t')}.
\label{eq:BogoDecay}
\end{multline}
\iac{where the effective drift coefficients $\bar{D}_k$ are determined by the shape of the Bogoliubov modes and tend to $D$ in the long-wavelength limit $k\to 0$. In this same limit, the lowest Bogoliubov mode has a diffusive character} with $\omega_k \simeq -i\gamma_k = -iJ^2\Gamma^{-1} k^4$ \cite{wouters2007}.
The factor $1/N_x$ in front of (\ref{eq:BogoDecay}) can be interpreted by viewing \iac{the} white spatial noise as randomly drawing noise realizations with a given $k$ and unit strength at each site. \iac{The} probability to pick \iac{a given mode} is then $1/N_x$. 

\iac{In a spatially finite system where $k$ is quantized, only the $k=0$ mode gives a finite contribution at long times, proportional to $|t|$; the contribution of all other modes decays instead exponentially with time. From this, one immediately obtains the expression of the Bogoliubov--Schawlow--Townes coherence time 
\begin{equation}
\tau_c=\tau_{ST} = 2 \gamma_{ST}^{-1} = \frac{2 n_0 N_x}{D}
\label{eq:Schawlow--Townes_xN}
\end{equation}
that generalizes the Schawlow-Townes phase diffusion to the case of a finite laser array\iaciac{, $n_0 N_x$ being equal to the total number of photons.}}

\iac{The situation is a bit different for an infinite array. In this case, the sum over discrete $k$ modes has to be replaced by an integral $k$. This yield~ the Bogoliubov prediction $g^{(1)}(t) \sim e^{-B|t|^{3/4}}$, where $B$ is a constant. The slower power-law decay stems from the fact that the specific $k=0$ mode is now occurring with a probability zero.}

\iac{For the sake of completeness, it is worth noting that a different scaling would be found } in the presence of a nonlinear refractive index\iac{. In this case, the imaginary part of the Bogoliubov frequency scales in fact as 
$\gamma_k \propto k^2$~\cite{wouters2007}, which leads to $g^{(1)}(t) \sim e^{-B|t|^{1/2}}$ for an infinite one-dimensional system.}

\subsubsection{Nonlinear KPZ effects}

\iac{All the results discussed in the previous subsection were based on a linearized Bogoliubov approximation where different modes are decoupled. Of course,  we know that this} approximation is not adequate for an infinite or large enough system, where \iac{nonlinear} KPZ features set in.
For an infinite  system \iac{a} stretched exponential behavior 
\iac{
\begin{equation}
 g^{(1)}(t) \sim e^{-B|t|^{2\beta}}
\end{equation}
was predicted in~}\citep{he2015}, with \iac{a universal} $2\beta = 2\chi/z= 2/3$ and \iac{a non--universal value of the constant $B$}.

If the system is \iac{sufficiently large but finite}, the Bogoliubov approximation \iac{breaks} down, but \iac{the spontaneously broken} $U(1)$ symmetry \iac{still imposes that the} coherence \iac{must decay at long times}  at least as fast as a pure exponential, $g^{(1)}(t) \sim e^{-|t|/\tau_c}$. \iac{In this case, we expect that the KPZ physics} typical of the infinite chain \iac{should remain} visible \iac{only for intermediate times}, up to a saturation time scaling as \iac{$(N_x)^z$}. 

\iac{These arguments on the functional form of the temporal decay of coherence are quantitatively illustrated
in Fig.~\ref{fig:crossover_KPZ_ST}, where we display the temporal} correlation function computed by \iac{numerically solving Eq.~(\ref{eq:array_resonators}) for three different system sizes} $N_x=64,256,1024$: \iac{the thick black line shows} $-\log g^{(1)}(t)$\iac{, while} the red and cyan lines are linear fits in the loglog scale of the plot.
Keeping the same observation window, for small \iac{system} sizes the \iac{temporal decay of the coherence $g^{(1)}(t)$} is mainly diffusive \iac{and follows an exponential law [panel (a)]. For larger sizes [panel (c)],} the \iac{exponential} Schawlow--Townes behavior is pushed at very long times \iac{so that only} the KPZ stretched exponential $g^{(1)}(t) \sim e^{-B|t|^{2/3}}$ is \iac{clearly visible in the time window displayed in the plot}.
\iac{An attempt to see the exponents of both regimes on a single plot is shown in the plot for an intermediate size shown in panel (b): while hint of them is visible, a complete separation of the two regimes would require a very large system} sizes and very long observation times, which is numerically \iac{very} demanding~\footnote{\iaciac{Note that the small downward deviations that are visible on the black curves next to the right edge of the plotting window (in particular in Fig.\ref{fig:crossover_KPZ_ST}.a) are a numerical artifact due to the large statistical spread of the late-time points. An explanation of its origin is given in Sec. SII of the SM \cite{suppl}.}}. 

\iac{The KPZ scaling of $g^{(1)}(t)$ at intermediate times is a clear indication of the crucial role of nonlinear coupling between modes in determining the phase dynamics. While the linearized Bogoliubov theory   predicts the (qualitatively correct) exponential \iaciac{form of the} decay of coherence at long times, it is natural to wonder whether the KPZ nonlinear couplings are responsible for any \iaciac{quantitative} deviation of the coherence time from the Bogoliubov--Schawlow--Townes prediction (\ref{eq:Schawlow--Townes_xN}).}

This issue is numerically investigated
~\footnote{
\iac{Note that the numerical effort required for these calculations grows up rapidly with $N_x$, since one has to access the dynamics at very long times, larger than the KPZ saturation time scaling as $\sim N_x^{3/2}$. For this reason, a statistical analysis of the errors has been restricted to the $N_x=128, J=5\gamma$ case, as expanded in Sec. SII of the SM \cite{suppl}}}
 in Fig.~\ref{fig:Nscaling} where we plot the numerical result for the coherence time $\tau_c$ in one-dimensional arrays of increasing sizes \iac{for two different values of the inter-site coupling $J=5\gamma$ (blue) and $J=0.5\gamma$ (cyan)}
.
\iac{ To better highlight the KPZ features, we have normalized the coherence time to} the single site Schawlow--Townes coherence time $\tau_{ST,1} \equiv \tau_{ST}/N_x = \frac{2 n_0}{D}$. \iac{For all parameter choices, the coherence time follows the Bogoliubov \iaciac{scaling proportional to $N_x$} until a certain critical size numerically compatible with the scaling  $l^*\sim J^{4/7}$, after which \iaciac{its increase with $N_x$ occurs} at a much slower rate.}

\iac{This marked deviation is indeed expected and can be understood looking at the  KPZ equation (\ref{eq:KPZ}): the total phase drift is the $\tilde{k}=0$ part of the phase field, which can be decomposed} in two statistically independent contributions
\begin{equation}
\tilde{\phi}(\tilde{x},\tilde{t}) = \tilde{\phi}_0(\tilde{t}) + \tilde{\phi}'(\tilde{x},\tilde{t}).
\end{equation}
\iac{Here,} $\tilde{\phi}_0$ \iac{accounts for the global phase evolution} generated by the $\tilde{k}=0$ component of noise, 
\begin{equation}
\partial_{\tilde{t}} \tilde{\phi}_0 \equiv \sqrt{\mathcal{D}} \xi_1(\tilde{k}=0, \tilde{t}),
\end{equation} 
and yields the Schawlow--Townes drift. \iac{Even though} the equation for $\tilde{\phi}'(\tilde{x},\tilde{t})$ is independent of $\xi_1(\tilde{k}=0, \tilde{t})$, an additional \iac{evolution of} $\tilde{\phi}'(\tilde{k}=0,\tilde{t})$ is \iac{induced by the finite $\tilde{k}$ components $\tilde{\phi}'(\tilde{k}\neq 0,\tilde{t})$ of the phase field via the KPZ nonlinearity $\lambda$. The additional phase noise induced by this nonlinear coupling is responsible for the deviation of the coherence time from the Bogoliubov-Schawlow-Townes prediction visible in Fig.~\ref{fig:Nscaling}. The complex behaviour of $\tau_c(N_x)$ shown in the figure suggests that a quantitative explanation of the phenomenon requires a non--perturbative analysis that will be the object of future work.}
\ivan{In particular, it is not even clear the exponent $\rho$ characterizing the large $N_x$  dependence $\tau_c \sim N_x^{\rho}$. A similar linewidth enhancement is expected also for other lattice dimensionalities, with $\rho$ smaller in lower dimensions, and for equilibrium atomic condensates at zero temperature.} 
\iac{For instance, Beliaev processes due to a nonlinear mode coupling were predicted in \citep{sinatra2009} to play a  role in the phase diffusion of equilibrium  condensates. \iaciac{Note however that the physical origin of the nonlinear mode coupling in atomic condensates is very different from the KPZ nonlinearity of lasers.}}
%


\section{Harper--Hofstadter topological laser}
\label{sec:topo}

As shown in recent theoretical \cite{harari2018, secli2019} and experimental  \cite{bahari2017, bandres2018} works, it is possible to make the edge mode of a \iac{photonic} topological insulator to lase by introducing a gain \iac{material into the device, while preserving the} topological properties such as the chirality of \iac{the edge mode} propagation and \iac{its topological} robustness in circumventing defects \iac{without suffering backscattering}.
\iac{This section is devoted to the study of the spatio-temporal coherence properties of such a device. Within the usual semi-classical approach, quantum and classical noise is described by including a white noise term to the equation of motion of the field. While the overall behaviour turns out to be very similar to the non-topological case studied in the previous Section, some interesting consequences of the topological nature of the mode are found and highlighted.}

\subsection{The model}

\begin{figure}[t]
\centering
\includegraphics[width=1.0\columnwidth]{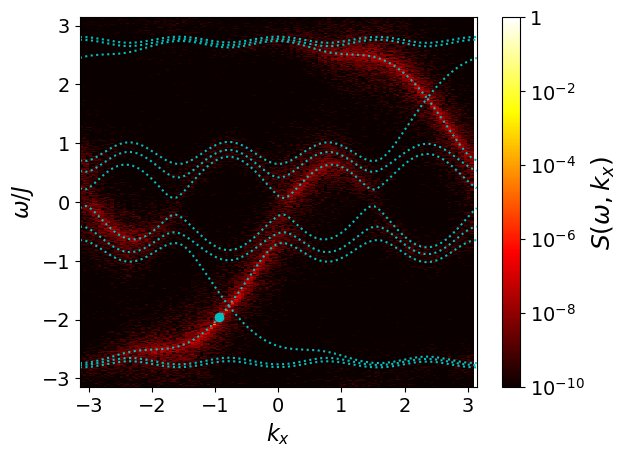}
\includegraphics[width=1.0\columnwidth]{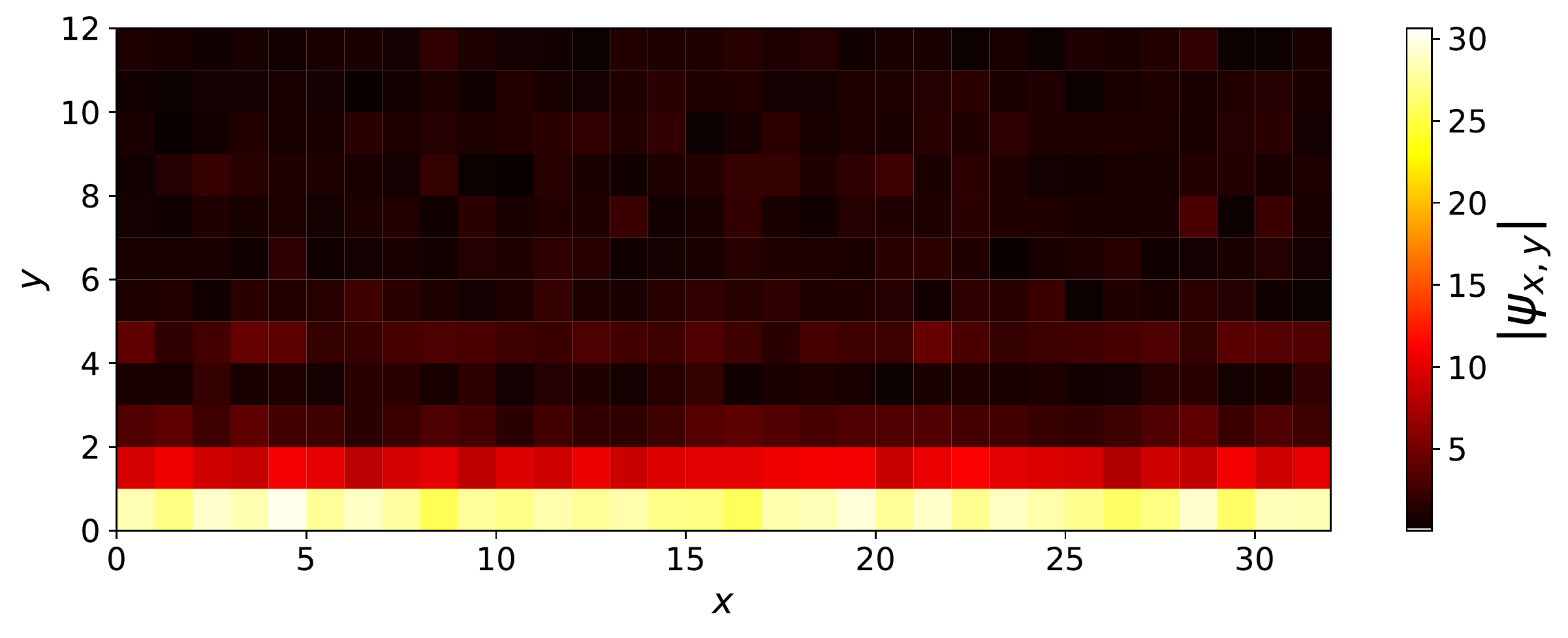}
\caption{{\it Topological lasing.} \iac{Lower panel: typical snapshot of the field modulus distribution $|\psi(x,y)|$ at steady-state. Upper panel: wavevector- and energy-resolved spectrum of the field on the $y=1$ edge.
The dotted lines are the Harper--Hofstadter bands and the spectral intensities are normalized to the laser emission at $\omega^{las} \simeq -9.774 \gamma$, $k_x^{las} = -2\pi\frac{19}{128}$.
Numerical calculations were performed according to Eq.~(\ref{eq:topo_laser}) for a lattice of size $N_x=128$, $N_y=12$ with periodic boundary conditions along $x$ and a flux density $\alpha=1/4$. System parameters: $J=5\gamma$, $P=2\gamma$, $n_s=1000$, $D_{x,y}=\gamma/2(1+\delta_{y,1})$. 
}}
\label{fig:TopoSpettro}
\end{figure}

\iac{With no loss of generality, we focus our attention on the topological bands of a two-dimensional Harper--Hofstadter model. This is a most celebrated topological model and is widely used in studies of topological photonics.}
By labeling with $x=1,..,N_x$ and $y=1,..,N_y$  the \iac{sites of the two-dimensional lattice}, the equations of motion for the field read
$$ 
i \partial_t \psi_{x,y} = - J \left[ \psi_{x,y+1} +\psi_{x,y-1} + e^{-2\pi i \alpha y}\psi_{x-1,y} + \right.
$$
\begin{equation}
\left. +e^{2\pi i \alpha y}\psi_{x+1,y} \right] + \frac{i}{2} \left[ \frac{P\delta_{y,1}}{1+ n_{x,y}/n_S}  - \gamma \right] \psi_{x,y} + \sqrt{2D_{x,y}} \xi_{x,y}
 \ , 
\label{eq:topo_laser}
\end{equation}
where $\alpha$ is the \iac{synthetic magnetic field flux per plaquette} in units of the magnetic flux quantum. \iac{As in many previous works, we focus on the $\alpha =1/4$ case. To simplify the geometry, we consider a} cylindrical lattice with periodic boundary conditions along the $x$ direction, and we introduce the gain medium on \iac{all} sites of the $y=1$ edge.
\iac{Inspired by the Wigner approach~\cite{gardiner2004}, we take the diffusion coefficient to have form $D_{x,y}=(1+\delta_{y,1})\,\gamma/2$. The stronger noise on the edge sites reflects the presence of gain. We have however checked that our results remain qualitatively identical if different spatial distributions of $D_{x,y}$ are used. We have also checked that the statistical results that are going to be discussed in the following of the paper are unchanged if different system geometries are considered, e.g. with open boundary conditions \iaciac{(see e.g. Fig. S3 of~\cite{suppl}).}}

\iac{For the sake of completeness, let us first} recall the properties of the \iac{underlying} Harper--Hofstadter Hamiltonian \iac{in the absence of gain, losses and noise.}
Within the Landau gauge used in Eq.~(\ref{eq:topo_laser}), implementing periodic boundary conditions along the $x$ direction makes the system translationally invariant, so that it is convenient to label states by the wavevector $k_x$. \iac{The dispersion for our $\alpha=1/4$ case is shown by the cyan dotted lines in Fig.~\ref{fig:TopoSpettro}: it consists of four bands of bulk states delocalized over the whole system, plus two chiral modes localized on each edge and energetically located in the gaps within the bands. On the $y=1$ side per example, the edge state in the lower (upper) gap has positive (negative) group velocity, and viceversa for the other side. The fact that there are no two counter--propagating states on the same edge at the same frequency is at the origin of the topological protection.}

\iac{We study the temporal evolution of the field according to the stochastic equations (\ref{eq:topo_laser}).} Starting the simulation with zero field, noise \iac{triggers laser operation by spontaneously breaking the $U(1)$ symmetry. A mean-field study of this physics in the absence of noise \iac{with} random initial conditions was presented  in \citep{secli2019}: 
at steady-state, quasi-monochromatic laser oscillation takes place in a mode which is randomly selected by the initial noise. The probability distribution is peaked at discrete frequencies roughly fixed by the eigenvalues of the underlying finite Harper-Hofstadter model. The distribution is symmetric with respect to zero frequency and has support in the energy gaps of the band structure. The maximum} corresponds to the eigenstates that are most localized on the edge for which the effective gain is the strongest: as it was pointed out in \citep{peano2016}, the $k_x$--dependent overlap of the Harper--Hofstadter eigenmode with the gain material determines a non-trivial dependence of the imaginary part of the dispersion. 

The main features of the \iac{steady-state} topological laser operation \iac{including noise} are illustrated in Fig.~\ref{fig:TopoSpettro}. In the lower panel we plot \iac{a typical example of} the field modulus $|\psi_{x,y}|$ for a \iac{finite} $N_x=32,N_y=12$ cylindrical lattice, showing localization of the mode on the edge. The upper panel reports \iac{instead} the power spectral density $S(k_x,\omega)$ of the field \iac{$\psi_{x,1}(t)$ on the $y=1$ edge: the narrow lasing mode is strongly saturated on this scale and is indicated by the cyan circle. Noise-induced fluctuations distribute themselves over all modes but are concentrated on the ones with largest overlap with the $y=1$ side, in particular on the two edge states with opposite chiralities. While the distribution roughly follows the dispersion of the modes in the underlying Harper--Hofstadter model indicated as a cyan dotted line, a complete theory requires a Bogoliubov analysis. This will be the subject of the forthcoming work~\citep{loirette-pelous2020}.}

\begin{figure*}[t]
\centering
\includegraphics[width=2.\columnwidth]{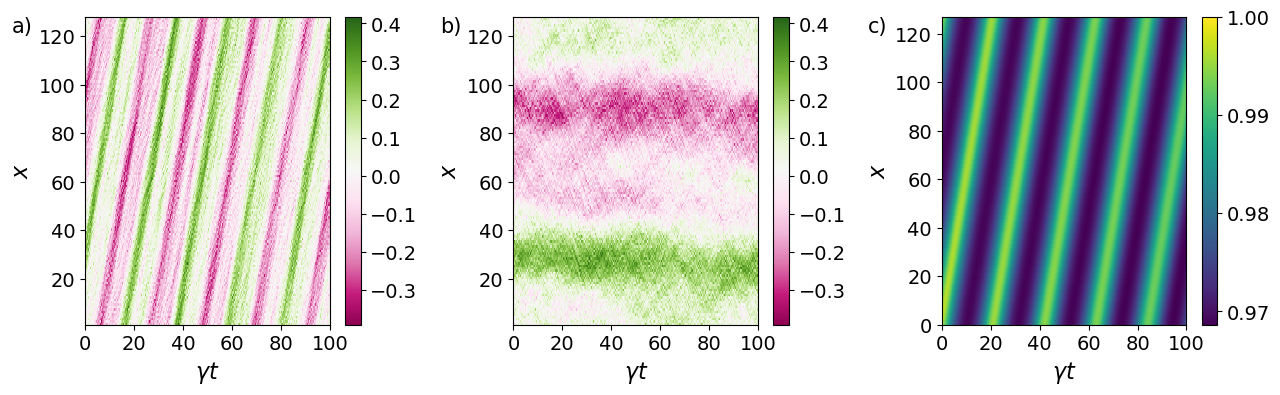}
\caption{\iac{{\it Chiral motion of the phase fluctuations}}. (a) \iac{Spatio-temporal plot of the slowly varying phase of} $\psi_{sl}(x,t)$, showing \iac{that the fluctuations chirally move around the system. At the steady state in the absence of noise, this plot would be flat.  (b) Spatio-temporal plot of the phase of $\psi_{_{CM}}(x,t)$ in the co-moving frame: now the fluctuations are observed in their natural frame of reference and evolve slowly.} (c) Correlation function $g^{(1)}(x, t)$: since the fluctuations move \iac{at $v_g$, after a time $t$} the field in $x$ is correlated with the \iac{one} in $x + v_g t$.
\iac{System parameters: $N_x=128, N_y=12, J=5\gamma, P=2\gamma, n_s=1000, D=2\gamma$. }}
\label{fig:to_comoving_frame}
\end{figure*}

\subsection{Chiral \iac{motion of the} phase fluctuations}

\iac{To characterize the spatio-temporal coherence of the emission, we consider the fluctuations of the phase of the one-dimensional field living on the amplifying boundary of the Harper--Hofstadter lattice, $\psi(x,t) \equiv \psi_{x,1}(t)$.}

\iac{In the steady-state of laser operation, the phase displays slow fluctuations around a carrier wavevector $k_x^{las}$ and frequency $\omega^{las}$: the former can be extracted from the (spatial) winding number of the phase around the system, the latter can be determined by fitting the evolution of the field phase on single sites. In the spectrum of Fig.\ref{fig:TopoSpettro}, they are indicated by the position of the cyan circle.
While a precise determination of these quantities can be important from the applicative point of view, they are somehow uninteresting from the statistical mechanics point of view, since they are mostly determined by the deterministic dynamics of the device and are weakly affected by the fluctuations. }

\iac{In order to remove the carrier frequency and wavevector and concentrate on the stochastic fluctuations, we define the slowly varying field 
\begin{equation}
{\psi}_{sl}(x,t) \equiv e^{\iac{-i(k_x^{las} x - \omega^{las}t)}}\,\psi(x,t).
\end{equation}
Looking at the phase of a typical realization of $\psi_{sl}(x,t)$ shown in Fig.~\ref{fig:to_comoving_frame}(a), we easily recognize a phase fluctuation pattern that moves at a constant velocity and gets slowly distorted. The drift velocity can be inferred from the dispersion $\omega_{em}(k_x)$ of the lasing chiral edge mode, which has  group velocity $v_g = \frac{d\omega_{em}}{dk_x}(k_x^{las})$  and curvature $J_{eff}(k_x^{las})= \frac{1}{2} \frac{d^2\omega_{em}}{dk^2_x}(k_x^{las})<J$. }

\iac{In order to focus on the intrinsic dynamics of the phase fluctuations, we plot in Fig.~\ref{fig:to_comoving_frame}(c) a typical realization of the phase evolution seen from the moving frame at $v_g$,
\begin{equation}
\psi_{{CM}}(x,t) \equiv \psi_{sl}(x+v_gt,t).
\end{equation}
For the relatively strong inter-site coupling $J=5\gamma$ and relatively small system size $N_x=128$, the phase fluctuations develop very slowly and remain quite small. Their magnitude gets larger if the mean intensity $n_0$ is reduced, the inter-site coupling $J$ is reduced, or larger systems are considered. This will be discussed in Sec.\ref{ssec:kpz}.}

While the transformation to $\psi_{_{CM}}(x,t)$ allows for a direct visualization of the phase dynamics, it is \iac{also} possible to study the fluctuations circulating along the edge by computing the space--time correlation function of the original field,
\begin{equation}
g^{(1)}(x, t) =
|\langle \psi^*(x ,t) \psi(0,0) \rangle|,
\end{equation}
where the average is \iac{taken over the} noise and invariance under \iac{temporal $t$ and spatial $x$} translations is assumed. \iac{This analysis requires no preliminary estimate} of $k_x^{las}$ \iac{and} $ \omega^{las}$ \iac{and will be our workhorse in the next sections.}
As \iac{it is} apparent looking at the smooth stripes in Fig.~\ref{fig:to_comoving_frame}\iac{(c), the analysis of $g^{(1)}(x,t)$ is the cleanest way to extract the velocity at which fluctuations travel. The result $v_g\simeq 6.07\, \gamma a$ is} perfectly compatible with \iac{the group velocity $v_g\simeq 6.08 \,\gamma a$ obtained from the linear dispersion of the chiral edge mode \iaciac{(see also Fig.~S4 of \cite{suppl})}.} 


\subsection{\iac{Correlated side-peaks in the emission spectrum}}

\begin{figure}[t]
\centering
\includegraphics[width=1.0\columnwidth]{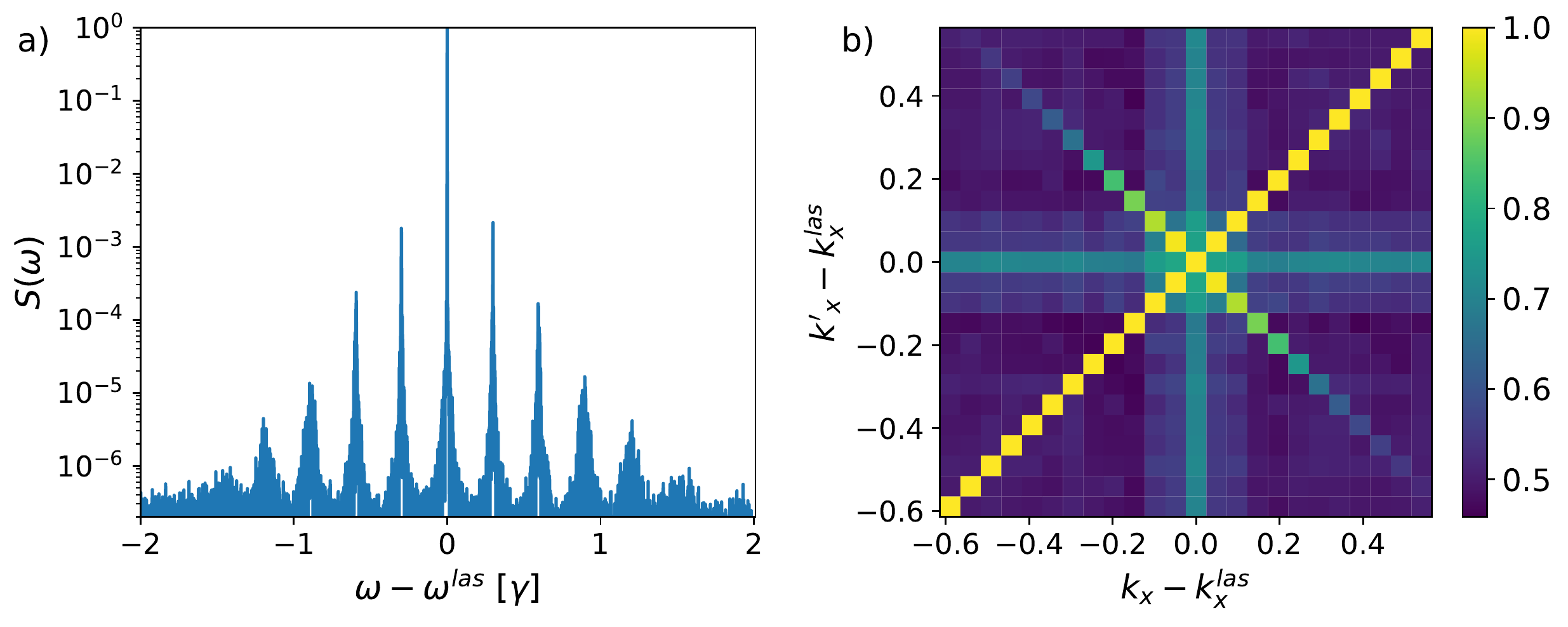}
\caption{\iac{{\it Correlated side-peaks in the emission spectrum.} (a) Spectrum of the field on a single resonator \iaciac{located} on the amplified edge. The spacing of the side--peaks is determined by the quantization of the wavevector $k_x$ around the periodic direction. The side--peaks are generated by parametric scattering processes from the lasing mode into pairs of symmetrically located modes. (b) Color plot of the momentum-space correlation function (\ref{eq:g2}) in the $k_x,k_x'$ plane showing -among other- strong correlations between symmetrically located modes. }}
\label{fig:comb}
\end{figure}

\iac{A complete understanding of the spectral distribution}  $|\psi(k_x, \omega)|^2$ \iac{obtained by spatio-temporal Fourier transform and} shown in Fig.~\ref{fig:TopoSpettro}(b) requires inspecting the Bogoliubov modes on top of the noiseless lasing state. This will be \iac{the subject of a forthcoming} work \citep{loirette-pelous2020}.

\iac{For the moment, we restrict our discussion to a few simple yet important remarks on the emission spectrum from each site.} In Fig.~\ref{fig:comb}(a) we show the \iac{emission spectrum defined as} 
\begin{equation}
S(\omega) = \frac{1}{N_x}\sum_x|\psi(x,\omega)|^2 
\end{equation}
for the parameters and lasing point shown in Fig.~\ref{fig:TopoSpettro}(b). \iac{In addition to the main lasing peak, the emission spectrum displays a comb-like structure with a series of symmetric side-peaks: the frequency spacing is determined by the quantization of the momentum along the periodic direction and is} approximatively $v_g \frac{2\pi}{N_x}$.

The visibility of the comb is not merely due to the existence of \iac{eigenstates at those specific values of the frequency, but their population by noise is enhanced by correlations. This is illustrated in} Fig.~\ref{fig:comb}(b) \iac{where we show} the normalized momentum--space intensity--intensity correlation function 
\begin{equation}
R^{(2)}(k_x, k'_x) = \frac{\langle n_{k_x} n_{k'_x} \rangle}{\sqrt{  \langle n_{k_x}^2 \rangle \langle  n_{k'_x}^2 \rangle   }}.
\label{eq:g2}
\end{equation}
\iac{Here, the} momentum--space densities \iac{$n_{k_x}$} are evaluated at the same time over \iac{the whole edge}, $n_{k_x}(t) = |\psi(k_x,t)|^2$. 

\iac{Several features are visible in this plot. The diagonal line for $k_x=k_x'$ is due to a trivial self-correlation and $R^{(2)}$ is here equal to $1$. For generic pairs of modes, the fact that $g^{(2)}(k_x)=2$ for $k_x\neq k_x^{las}$ implies that the background value is $R^{(2)}=0.5$. When one (two) modes coincide with the lasing one, $R^{(2)}$ is equal to $1/\sqrt{2}$ (1), which explains the vertical and horizontal stripes and the central peak.}
\iac{The most interesting feature is the stripe on the anti-diagonal, corresponding to correlations between symmetrically located modes such that $k_x+k_x'=2 k_x^{las}$. For the first two pairs of side--peaks, this correlation is nearly perfect, indicating that these modes are populated in pairs by parametric scattering processes.}

\iac{Such correlations are of course not specific of topological systems but can be observed also in the trivial systems studied in the previous Sec.~\ref{sec:trivial}, albeit with a suppressed intensity due to the curvature of the dispersion. In analogy to exciton-polariton systems pumped around the magic angle~\cite{carusotto2013}, the magnitude of these parametric correlations would be reinforced again if lasing was made to operate around the inflection point of the dispersion.}

\subsection{KPZ \iac{spatio-temporal} correlations}
\label{ssec:kpz}

\begin{figure}[t]
\centering
\includegraphics[width=1.\columnwidth]{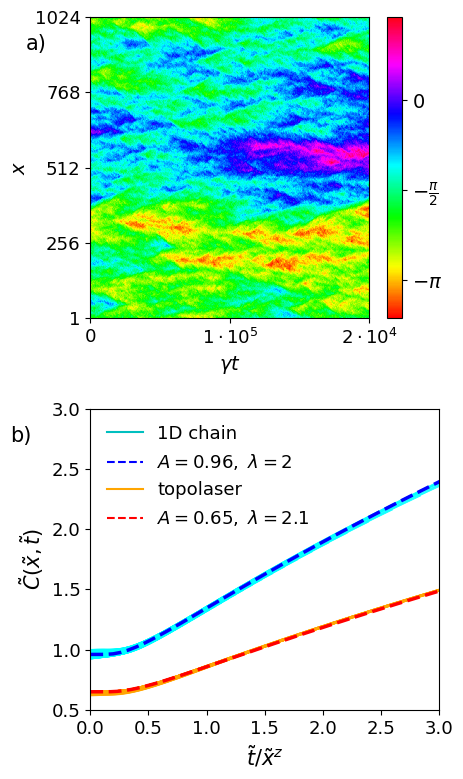}
\caption{\iac{{\it KPZ dynamics of topological lasing}. (a) Typical example of the space-time dynamics of the phase of the field $\psi_{_{CM}}(x,t)$ on the system edge seen from the comoving frame. (b) Correlators $\tilde{C}(\tilde{t}, \tilde{x}^z)$ for $x=\pm 30,...,\pm 160$ (small lines) for the topological device lasing into the $k_x=-2\pi\frac{155}{1024}$ mode (orange) and for the non-topological one-dimensional array lasing in the $k_x=0$ mode (cyan). Red and blue dashed lines indicate the KPZ universal function (\ref{eq:KPZcorrelators}) on which all curves collapse. System parameters: $N_x=1024$, $J=0.5\gamma$, $P=2\gamma$, $n_S=1000$, $D_{x,y}=\gamma/2(1+\delta_{y,1})$.}}
\label{fig:KPZ_scaling}
\end{figure}

\begin{figure*}[t]
\centering
\includegraphics[width=1\textwidth]{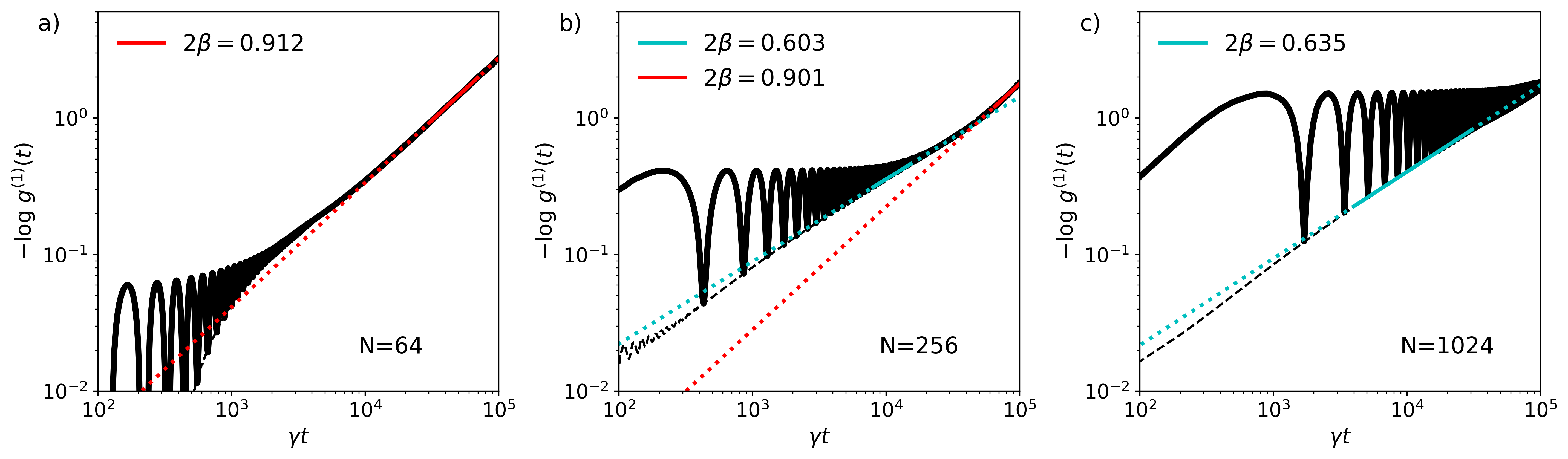}
\caption{\iac{{\it Kardar-Parisi-Zhang to Schawlow-Townes crossover in a finite--size topological laser.}}  
This plot is the analogous of Fig.~\ref{fig:crossover_KPZ_ST} for the topological laser. The thick black lines correspond to the logarithm of the temporal correlation for a given site on the edge of the lattice. \iac{The thin dashed lines show the temporal evolution of the 
coherence $-\log g_{_{CM}}^{(1)}(t)$ seen from the reference frame comoving with the edge state.}
For increasing system sizes and a given temporal window, a crossover between an \iac{exponential} (red fits) and \iac{a} KPZ (cyan fits) decay of the coherence is observed. The amplitude of the oscillations is inversely proportional to the spatial coherence of the device. \iac{In the long-time limit}, only the global phase matters and the oscillations fade away.  
}
\label{fig:topo_crossover_KPZ_ST}
\end{figure*}

\iac{The discussion of the emission spectrum presented in the previous subsection gives first hints of the complexity of the multi-mode dynamics of fluctuation dynamics. Here we will build a complete theoretical picture} of the spatio-temporal \iac{coherence} of the topological laser. 

\iac{To get a first hint on the behaviour, we can make} the assumption that the field on the edge can be described in the comoving frame by the 1D equation
\begin{multline}
i\partial_t \psi_{_{CM}}(x,t) =  (-J_{eff} + i \eta)\nabla^2 \psi_{_{CM}} +  \\ + \frac{i}{2} \left[  \frac{P_{eff}}{1+n_S |\psi_{_{CM}}|^2} - \gamma \right] \psi_{_{CM}} + \sqrt{2D}\, \xi.
\label{eq:etaKPZ}
\end{multline}
Here $J_{eff}$ is given by the curvature of the bare Harper--Hofstadter topological mode, $P_{eff}$ is chosen as to retrieve the numerical mean intensity $n_0$ on the edge, and $\eta$ accounts phenomenologically for the \iaciac{$k$}-dependent localization of the lasing mode \iac{on the edge of the lattice}. A \iac{rigorous} account of this dimensional reduction \iac{procedure} will be presented in \citep{loirette-pelous2020}.

\iac{Assuming a fast relaxation of the intensity fluctuations, we can then restrict our attention to the dynamics of the phase.} By neglecting terms containing four derivatives (both the linear, Galilean-preserving ones and \iac{the} nonlinear, Galilean-breaking \iac{ones} as shown in  Sec. SI~\cite{suppl}), one gets \iac{to a motion} equation for the phase \iac{of the KPZ form}:
\begin{equation}
\partial_t \phi =  \eta \nabla^2 \phi +J_{eff} (\nabla \phi)^2 +  \sqrt{\frac{D}{n_0}} \xi.
\end{equation}
Since $\eta$ is the \iac{less controlled} parameter of the model, we do not perform the usual KPZ rescaling \iac{\citep{he2015}} to yield an equation containing the effective nonlinearity as the only one  parameter. Rather, we rely on the rescaling Eqs.~(\ref{eq:star_units}) with the effective parameters $J\to J_{eff}, P\to P_{eff}$; this transformation does not depend on $\eta$ \footnote{Actually, also the dependence on $\Gamma_{eff}$ is such that, given the measured $n_0$, any uncertainty on the value of $P_{eff}$ will only affect the KPZ coefficient of the Laplacian.} and yields
\begin{equation}
\partial_{\tilde{t}} \tilde{\phi} = \frac{\eta t_*}{l_*^2} \tilde{\nabla}^2 \tilde{\phi} + (\tilde{\nabla} \tilde{\phi})^2 + \xi.
\end{equation}
\iac{In} these units we  then expect the KPZ nonlinearity to be close to $\lambda=2$; this value is also protected from renormalization induced by the quartic derivative \iac{term in the KSE} \citep{ueno2005}. \iac{It is thus natural to expect that the coherence of the topological laser will closely resemble the one of the generic extended laser discussed in Sec.\ref{sec:trivial}. In what follow, we proceed to numerically verify this statement on simulations of the stochastic laser equations in the topological two-dimensional lattice.} 

\iac{Based on our previous discussion, we expect that the} KPZ universal dynamics occurs, in a lattice of $N_x$ sites, on timescales shorter than the saturation time $\sim N_x^z$ \iaciac{(}after which the Schawlow--Townes like behavior described above sets in\iaciac{)} but  larger than the  timescales where \iac{the linear Bogoliubov dynamics} and non--universal \iac{effects} dominate. \iac{Having a sizable window where to observe KPZ physics then requires the system to be large enough, precisely it should be at least $N_x a \gg \sqrt{2}\pi l^*$ \citep{gladilin2014}.}
\iac{We thus} consider a \iac{large} system of length $N_x = 1024$ \iac{with periodic boundary conditions along $x$ and with $N_y=12$ points along the open direction $y$. In order to clearly observe KPZ physics while keeping intensity fluctuations within 15\% and having a tractable system size, it is beneficial to use a small inter-site coupling $J=0.5\gamma$.}

\iac{One may argue} that \iac{such a value of the coupling} $J$ \iac{(and thus of the topological gap)} is comparable with the bare linewidth $\gamma$. Such narrow topological gaps are very relevant for experimental implementations\cite{bahari2017}, but it is not \iac{a priori} obvious \iac{whether in this regime the} chiral edge modes \iac{survive losses. While this is indeed a serious issue to observe chiral edge propagation in passive systems, it is a crucial result of laser theory that the laser linewidth above threshold can be orders of magnitude smaller than $\gamma$. Our numerical simulations confirm stable lasing into the chiral edge mode even for for small $J/\gamma$; in particular, the robustness of topological lasing has been verified in the presence of a strong defect on the edge, as reported in Fig. S5 \cite{suppl}. }

To \iac{numerically highlight the KPZ physics} we have performed 20  simulations of \iac{the full two-dimensional lattice of} duration $\gamma T=5\cdot10^5$, starting from \iac{a plane wave with the wavevector value} for which the Harper--Hofstadter eigenstate is most localized on $y=1$, that is $k_x=-2\pi\frac{155}{1024}$.
For each run the correlation function  
$\langle \psi^*(x ,t) \psi(0,0) \rangle$ \iac{on the edge site} is computed and then averaged over the 20 trials to yield $g^{(1)}(x, t)$.  
The typical dynamics occurring in a time $\gamma T=2\cdot10^4$ is depicted in Fig.~\ref{fig:KPZ_scaling}(a), where the phase of the field $\psi_{_{CM}}(x,t)$ along the edge is shown \iac{in the comoving frame: a structure similar to the fractal structure} of interface growth can be recognized. 

Defining the correlation function in the comoving frame \iac{as}
\begin{equation}
 g_{_{CM}}^{(1)}(x,t) = |\langle \psi_{_{CM}}^*(x ,t) \psi_{_{CM}}(0,0) \rangle|
\end{equation}
 and the rescaled correlator \iac{as}
\begin{equation}
\tilde{C}(\tilde{t}, \tilde{x}^z) \equiv -2 (\phi^*)^{-2}  \tilde{x}^{-2\chi}  \log g_{_{CM}}^{(1)}(\tilde{x},\tilde{t}),
\end{equation}
 KPZ universality requires \iac{that}
\begin{equation}
\tilde{C}(\tilde{x}, \tilde{t}) = 
  \tilde{C}(\tilde{t}/\tilde{x}^z)
  = \left( \frac{1}{2}\lambda A^2 s \right)^{2/3} g_{KPZ}\left(\frac{1}{(2\lambda^2 A s^2)^{1/3}} \right)
  \label{eq:KPZcorrelators}
\end{equation}
with $s = \tilde{t}/\tilde{x}^z$ \iac{and $z=3/2$} on the right-hand side \iac{and $g_{KPZ}$ a universal function discussed in Sec.\ref{ssec:trivial_g1}.}
In particular, one has $\tilde{C}(0)=A$. 

 In Fig.\ref{fig:KPZ_scaling}(b) we plot  $\tilde{C}(\tilde{x}, \tilde{t})$ for $x=\pm 30,...,\pm 160$ and demonstrate that \iac{within an} excellent approximation they \iac{only} depend on  $\tilde{t}/\tilde{x}^z$ (to correct for the finite size effects, we actually plot $\tilde{C}(\tilde{x}, \tilde{t}) / (1-|x|/N_x)$, as illustrated in Fig. S6 \citep{suppl}).
 The  orange and cyan series correspond \iac{to calculations for} respectively the topological laser and the \iac{trivial one-dimensional chain laser in the $k=0$ mode discussed in Sec.\ref{ssec:trivial_g1}. The same physical parameters have been used in both cases}. 
 
The crucial point about Fig.\ref{fig:KPZ_scaling}(b) is that \iac{the phase, the space and the time have} been rescaled by \iac{the $\phi^*$, $l^*$, $t^*$ values obtained by using the effective} masses and gain parameters: for the topological laser, $J_{eff}$ is the curvature of the Harper--Hofstadter band at the lasing point and $P_{eff}$ is chosen \iac{in order to reproduce the observed intensity on the edge.}
For instance, $J_{eff} \simeq 0.319\gamma$ hence $l^* \simeq 1.92$ here. A detailed report of possible mappings of the $2D$ topological laser to a $1D$ one will be presented soon\citep{loirette-pelous2020}.

As anticipated, the Renormalization Group analysis \citep{ueno2005} predicts for the $1D$ array lasing in the $k^{las}=0$ mode that the rescaled KSE Eq.~(\ref{eq:KPZ}) flows to the low energy effective KPZ theory Eq.~(\ref{eq:KSE_tilde})
with $\lambda=2$, since, thanks to the Galilean invariance holding for KSE and KPZ equations, the nonlinear coupling is not renormalized.
Indeed, in this case the rescaled correlation functions  collapse to a unique curve \iac{as shown in Fig.\ref{fig:KPZ_scaling}(b) and this curve is excellently fitted (blue dashed line) using (\ref{eq:KPZcorrelators}) with $\lambda=2$ and $A=0.96$, as expected from the Galilean invariance argument. }
For the topological laser, \iac{the curves again collapse onto a single curve, which is well fitted using $\lambda=2.1 $ and $A=0.65$ (red dashed line). The value $\lambda=2.3$ with $A=0.63$ is an upper bound for the fitted value of the nonlinearity.}

Notice that for the topological laser there is no \iac{a priori} guarantee that a Galilean invariant KSE holds microscopically; 
on the contrary, an analysis on the lines of \citep{gladilin2014,altman2015} suggests that \iac{a} rescaling with $J_{eff}$ still yields a  microscopic $\lambda=2$, \iac{but other terms should also be added in Eq.~(\ref{eq:KSE_tilde}), e.g. of the kind 
$\nabla^2 \phi (\nabla \phi)^2$, $(\nabla^2 \phi)^2$, $\nabla^3 \phi \nabla \phi$, etc.}
These additional terms come from effective imaginary derivatives due to the $k_x$ dependent localization of the Harper--Hofstadter eigenstates on the edge; in particular, they break Galilean invariance and \iac{one may expect that they significantly renormalize} the effective KPZ parameters, since \iac{they contain the same number (four) of derivatives as the KSE term. From the numerics, it turns out that the renormalization of $\lambda$ is instead small.}

\iac{Still, it is interesting to note that the curves for the topological laser (in the properly rescaled units) sit below the ones of the trivial one-dimensional laser array.}
This \iac{feature can be traced back to} the imaginary term \iac{proportional to $\eta$ in (\ref{eq:etaKPZ}) that accounts for the $k$-dependence of the edge mode penetration in the bulk. This term} stabilizes \iac{the emission} and makes the topological device more coherent than the 1D laser with \iac{the corresponding} $J_{eff}, P_{eff}$. The ``ultraslow decay of fluctuations'' that was observed in \citep{secli2019} is indeed a consequence of the Goldstone branch and thus is a general feature of spatially extended lasers. The crucial role of $\eta$ in topological devices is apparent already at the level of Bogoliubov analysis \citep{loirette-pelous2020}, at least for class-A lasers. \iac{It is also} remarkable \iac{that the very noisy field in the bulk \iaciac{that is visible e.g. Fig. S2 of \cite{suppl}}} does not impact the coherence of the edge mode.

\iac{We conclude this section with} a brief remark on the experimental protocol to assess KPZ physics. The analysis of the correlation functions $g_{_{CM}}^{(1)}(x, t)$ \iac{shown in} Fig.\ref{fig:KPZ_scaling}.b \iac{was} carried \iac{out} in the \iac{reference} frame comoving with the chiral mode,  while in the lab one can easily measure correlation functions between different points at different times $g^{(1)}(x, t)$. \iac{However, since the two correlations functions are simply related by $g_{_{CM}}^{(1)}(x, t) = g^{(1)}(x-v_g t, t)$, the interesting $g_{_{CM}}^{(1)}$ can be extracted by a straightforward post-processing of  $g^{(1)}(x, t)$. Graphically,} this amounts to 
\iac{tilt the correlation function of Fig.~\ref{fig:to_comoving_frame}(c) with the suitable $v_g$} so to have the maximum of $g^{(1)}(x, t)$ \iac{at $x=0$ for all times $t$ (see Fig. S4 \cite{suppl}).}

\subsection{Temporal coherence and linewidth}
\label{ssec:topo_line}

\iac{After having confirmed that the long-time, large-distance spatio-temporal coherence of the topological laser is well described by the KPZ model, we now proceed to investigate the problem of the coherence time of a realistic, finite-size device.
Rather than analyzing with high resolution the linewidth of the main spectral peak of Fig.~\ref{fig:comb}(a), we work in real time and, with a concrete optical experiment in mind, we monitor the temporal coherence of the emission from a given site. 

\subsubsection{\iaciac{KPZ to Schawlow-Townes crossover}}

As compared to our discussion of the trivial case in Sec.\ref{ssec:trivial_g1}, there is the complication that phase fluctuations undergo a chiral motion around the system, so the coherence function of a given site displays the strong \iaciac{temporal} oscillations \iaciac{visible} in Fig.~\ref{fig:to_comoving_frame}(c).}



\iac{As a first step, we illustrate the crossover between  KPZ  coherence decay and Schawlow--Townes for different system sizes $N_x$. In Fig.~\ref{fig:topo_crossover_KPZ_ST} we show the temporal evolution of the phase diffusion $-\log g^{(1)}(t)$ \iaciac{(thick solid black lines)} and $-\log g_{_{CM}}^{(1)}(t)\equiv -\log g_{_{CM}}^{(1)}(0,t)$ \iaciac{(thin dashed  black lines)} in a given temporal window for increasing $N_x=64,256,1024$. 
As expected, these curves show sharp local minima of $-\log g^{(1)}(t)$ corresponding to local maxima of the coherence. As they originate from chiral motion of fluctuations around the system, these oscillations have a period $N_x/v_g$. The value of the coherence at these minima provides a discrete sampling of the equal-space coherence function $g_{_{CM}}^{(1)}(t)$. Similarly, the local maxima (minima of coherence) provide a sampling of  $g_{_{CM}}^{(1)}(N_x/2,t)$.}

\iac{Looking at the envelope of the minima for the largest system, we see that the agreement of Fig.~\ref{fig:topo_crossover_KPZ_ST}(c) with the KPZ result is very good. In particular, it makes a clear distinction from the predictions of the linear Edwards--Wilkinson model for which the exponent would be different.
Note also that the oscillations in $-\log g^{(1)}(t)$ are well visible at short times but fade away at long times, where only the global phase of the field over the whole lattice  matters and no oscillation is any longer visible. For the same reason, monitoring a resonator not on the edge one sees a  fast initial decay, but at long times the phase diffusion is the same as the edge, as sketched in Fig. S2 \citep{suppl}. }

\iac{For the smaller systems, the different functional form shown in Fig.\ref{fig:topo_crossover_KPZ_ST}(a) displays an exponential decay of the late-time coherence. As one can see comparing with panel (b), the late-time exponential decay is always there, it is just pushed to extremely long times in the largest systems. This is in close analogy with what we had found in Sec.\ref{subsec:linewidth_trivial} for the trivial system.}

\subsubsection{\iaciac{Linearized Bogoliubov prediction for the linewidth}}

\iac{The natural question is now to estimate the rate of this exponential decay and see how it scales with the size of the system. As a first step, we adopt a linearized Bogoliubov approach. For generic systems of $N$ sites (labelled as $\vec{x}$) of arbitrary dimensionality}, let us call $\mathcal{L}_{las}$ the $2N\times 2N$ Bogoliubov matrix of the linearized dynamics on top of the lasing steady-state. 
Let $V=\{  V_{\vec{x}\sigma,p} \}$ be the invertible matrix which diagonalizes $\mathcal{L}_{las}$, where \iac{the pseudo-spin} $\sigma =\uparrow,\downarrow$  \iac{indicates} the particle and hole components of the Bogoliubov problem and $p$ labels the $2N$ eigenmodes.
The Goldstone mode $V_{\vec{x}\sigma,G}$, that we assume to be unique with all other excitations having a finite life-time, is the eigenstate with zero eigenvalue. \iac{As usual, its spatial shape follows the one of the lasing mode.} 
The effective noise acting on the lasing mode will be determined by the projection of the bare noise on \iac{it. For} generality we consider a position dependent bare noise $D_{\vec{x}}$. Then, in the Bogoliubov approximation, the phase drift associated with the Goldstone mode \iac{is given} by
\begin{equation}
\langle \Delta_t \phi^2 \rangle =  \Phi_G^2 \sum_{\vec{x}} D_{\vec{x}} \ (V^{-1})_{G,\vec{x}\uparrow} (V^{-1})_{G,\vec{x}\downarrow} \ |t|
\label{eq:BogoDrift}
\end{equation}
where the summation represents the projection of the noise on the Goldstone mode and
\begin{equation}
\Phi_G = \Phi_G(\vec{x}) = \frac{-i}{n_0(\vec{x})}\left[ \psi^*_0(\vec{x})V_{\vec{x}\uparrow,G} - \psi_0(\vec{x})V_{\vec{x}\downarrow,G}\right]
\end{equation}
is actually independent of $\vec{x}$ \iac{and provides the normalization of the Goldstone mode phase component.}

\iac{If $VV^{\dagger} = 1$ a very clear expression holds for the Schawlow--Townes line:
\begin{equation}
\gamma_{ST} = \frac{\sum_{\vec{x}} D_{\vec{x}} n_{\vec{x}} }{ n^2_{tot}  } = \frac{D_n}{n_{tot}}
\end{equation}
with $n_{tot} = \sum_{\vec{x}}  n_{\vec{x}}$ and  $D_n = \sum_{\vec{x}} D_{\vec{x}} n_{\vec{x}} / \sum_{\vec{x}}  n_{\vec{x}}$.
\iac{This is for instance the case of} a spatially uniform, topologically trivial system, for which the different wavevectors decouple and the sector of $\mathcal{L}_{las}$ at the lasing wavevector $\vec{k}_{las}$ is diagonalized by a $2\times 2$ unitary matrix  $VV^{\dagger} = 1$, since modulus and phase are decoupled for the lasing mode.}

\iac{The situation is a bit more complicated in the topological case: the $k_x=0$ sector has dimension $2N_y\times 2N_y$ and $V$ is not unitary $VV^{\dagger} \neq 1$, so the theory discussed above does not hold. However, for the considered parameters it turns out that $VV^{\dagger} \simeq 1$, so the approximate expression
\begin{equation}
\tau_{ST} = \frac{2}{\gamma_{ST}} \simeq 2  \frac{\left[ \sum_{x,y} n_{x,y} \right]^2}{\sum_{x,y} n_{x,y} D_{x,y}} = 2 N_x  \frac{\sum_y n_{1,y}}{D_n}.
\label{eq:BogoST}
\end{equation} 
is expected to provide an accurate approximation.}
In \iac{the} standard laser theory the non--orthogonality of $V$ is also known under the terminology of Petermann factor and excess noise  \cite{siegman1989a,siegman1989b}. \ivan{The transverse Petermann factor, defined here as the ratio $\mathcal{K}=\frac{\gamma_{ST}}{D_n/n_{tot}}$, for the topological laser device is around $\mathcal{K} \simeq 1.002$ for $J=5\gamma$ and $\mathcal{K} \simeq 1.1$ for $J=0.5\gamma$ (the difference to be attributed to gain guiding), meaning that\iaciac{, within the linearized Bogoliubov approximation,} the laser emission is 
\iaciac{for} all practical purposes determined only by the  total number of photons in the device, which is the textbook, optimal case. }

\subsubsection{\iaciac{Numerical results for the linewidth}}

\iaciac{In order to assess the validity range of the Bogoliubov calculation, numerical simulations of the stochastic equations have been performed for the full two-dimensional model. The numerical predictions for the coherence time of the topological laser are} \iac{shown by the red and orange triangles in Fig. \ref{fig:Nscaling} as a function of the system size. The dashed line shows the theoretical prediction Eq.~(\ref{eq:BogoDrift}). From these results, one concludes that the topological laser  behaves again similarly to the topologically trivial one-dimensional laser array: for small $N_x$ the agreement with the Bogoliubov-Schawlow-Townes model of phase diffusion is excellent and the coherence grows proportionally to $N_x$. On the other hand, marked deviation occur for larger systems with a slower-growing coherence time. (Notice that the dependence of the Petermann factor on $N_x$ is negligible and via $k_x^{las}$.) }

\iac{To conclude our discussion, it is interesting to note that \iaciac{the Bogoliubov-Schawlow--Townes prediction that well captures the emission linewidth for small systems does not depend on $J_{eff}$ nor on the dispersion of the Bogoliubov modes at $k_x\neq 0$.}} On the other hand, the deviation observed for larger systems does strongly depend on $J_{eff}$\iaciac{, which} pinpoints the crucial role of the  KPZ nonlinearities illustrated above~\footnote{\iac{For this plot we chose the $k_x^{las}$ corresponding to the maximally localized Harper--Hofstadter eigenvector, but the results are qualitatively independent of this choice. Fixing $k_x^{las}$ is however needed if one is to compute the KPZ correlation functions by running parallel simulations.}}.

\section{Lasing with on--site disorder}
\label{sec:disorder}

\iac{The general message of the previous Section was that the coherence properties of a topological laser follow the same KPZ dynamics as the ones of a topologically trivial, one-dimensional laser array. This conclusion is not restricted to the well-known KPZ features in the infinite system limit, but also applies to the dependence of the coherence time on the system size and to the marked deviations from the \iaciac{single-mode} Schawlow-Townes prediction.}

\iac{In this Section we investigate the effect of static disorder on the coherence of the laser emission. A certain degree of fabrication imperfections and inhomogeneities is in fact expected to be always present in realistic devices. Our numerical study points out a dramatically different behaviour of \iaciac{topologically} trivial vs. topological systems: disorder has a strong impact on the coherence of a topologically trivial system, a small amount of disorder being able to give a wide range of realization-dependent, chaotic and multi-mode phenomena. On the other hand, the temporal coherence of a topological laser is robust against a sizable disorder and emission remains well monochromatic as long as the disorder magnitude is not so large to close the topological gap.}


\begin{figure}[t]
\centering
\includegraphics[width=1.0\columnwidth]{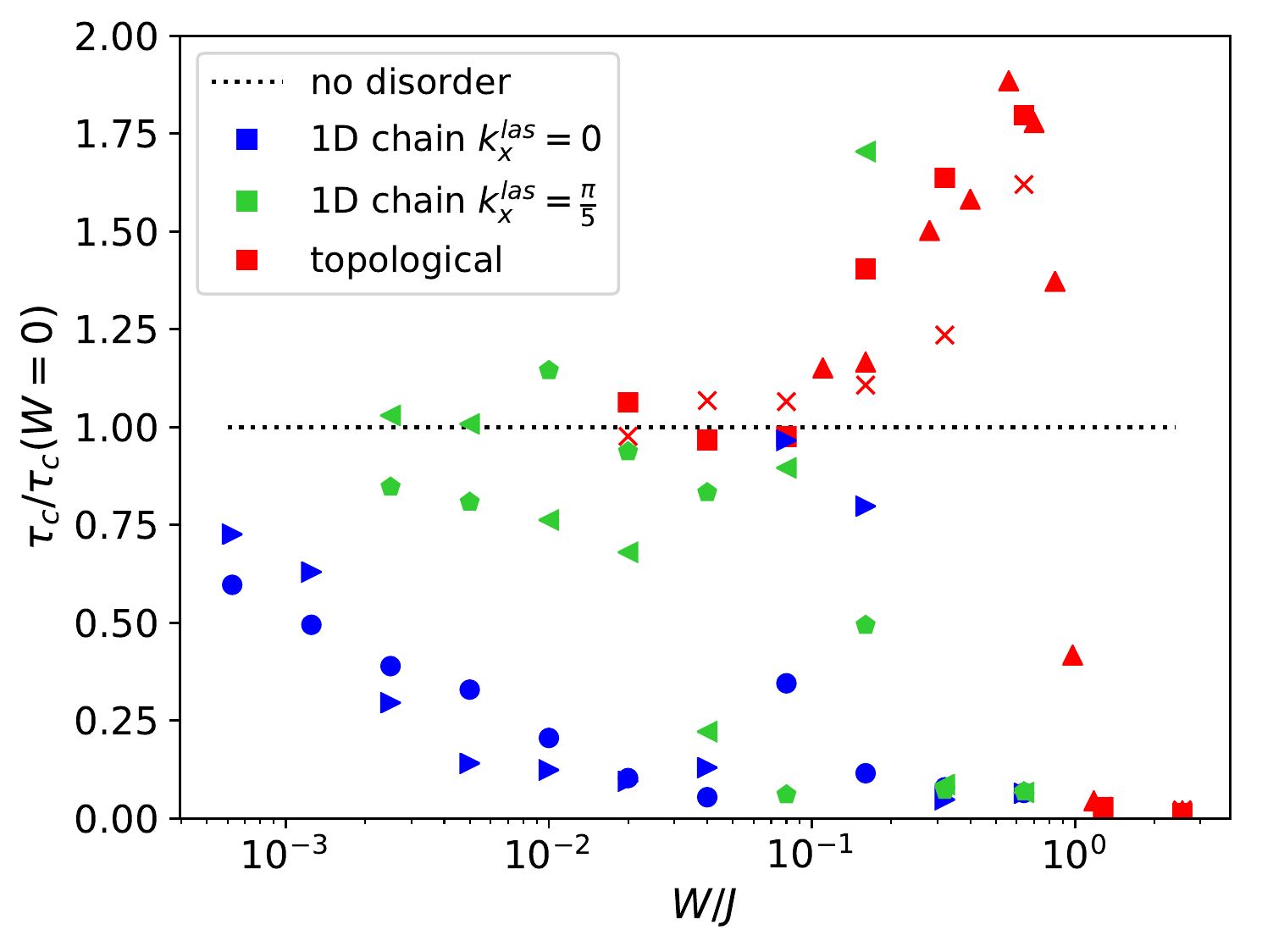}
\caption{ \iac{{\it Topological robustness of the temporal coherence.} Plot of the coherence time (normalized to the clean system value) as a function of the strength $W$ of the disorder. Different markers refer to different realizations of disorder. \iaciac{Blue and green} markers are for a non-topological one-dimensional laser array lasing at $k^{las}=0$ (blue) or $k^{las}=\frac{\pi}{5}$ (green). Red markers are for the topological laser with periodic boundary conditions (squares and triangles) and with open boundary conditions (crosses). Same marker shapes correspond to the same spatial distribution of the disorder potential except for the overall strength $W$.}  }
\label{fig:tau_W}
\end{figure}
%

\subsection{Lasing in a non-topological one-dimensional array with disorder}

\iac{We start by considering  the effect of on--site disorder on the lasing properties of a non-topological array of resonators. We do not aim here to a general discussion of the theory of lasing in disordered systems or to make a connection with random lasers~\citep{wiersma2008}, but our purpose is just to provide a benchmark to assess the features of a topological laser.}


Along all \iac{this section}, a disordered potential is added to Eqs.~(\ref{eq:array_resonators}) and (\ref{eq:topo_laser}) \iac{in the form},
\begin{equation}
i\partial_t \psi_{\vec{x}}  =  \iac{\ldots +} W \mathcal{G}(0,1) \psi_{\vec{x}},
\end{equation}
where $\mathcal{G}(0,1)$ is a Gaussian  random variable with mean 0 and variance 1. \iac{For the sake of definiteness,} we restrict \iac{here} to \iac{the} $N_x=128$ and $J=5\gamma$ \iac{case}.
The lasing dynamics in the presence of disorder is in general very complex, but, since our ultimate goal is a qualitative comparison with the topological laser, we focus here on the coherence time of the system for various values of disorder $W$, and in particular on \iac{looking} whether there is a \iac{clear} threshold \iac{value of disorder above which} coherence collapses.

For linear waves, the sensitivity of the eigenstates at a given energy to a static perturbation is proportional to the spectral density of states.
Then, in order to have a \iac{fair comparison of the trivial and topological cases}, we consider \iac{lasing both at $k^{las}=0$ and at $k^{las}=\frac{\pi}{5}$. This latter case has a finite group velocity (and hence a density of states) comparable to the one of the chiral edge mode of the Harper--Hofstadter model in its central part} \ivan{and for these reasons it seems the most natural way of benchmarking the topological laser.}

\iac{In Fig.~\ref{fig:tau_W}, we plot the coherence time normalized to the value in the absence of disorder. In particular, we use an exponential fit to extract the coherence time $\tau_c$ for each site (an exponential fit is used even if the shape of $g^{(1)}(t)$ is in general very complex) and we plot the average over the lattice.
Markers with the same shape indicate that the same realization of disorder and the same initial conditions have been used, while only \iaciac{the overall strength factor} $W$ is varied.}

\iac{From the plot, we see that already a very small disorder has a marked impact on the coherence time of the device, the threshold \iaciac{value} depending on $k_x^{las}$. While a detailed description of all possible behaviours depending on the disorder realization goes beyond the scope of this work (some illustrative examples are shown in the SI \cite{suppl}) and not fully understood non-monotonic behaviours are observed in some cases, we can safely conclude that the coherence is quickly lost in most of the realizations. For strong disorder $W \sim 0.8 \gamma$ laser operation gets fragmented and different portions of the sample end up lasing at different frequencies (see Fig. S7 of \cite{suppl}).}

\subsection{Lasing in a topological disordered system}

\iac{The same protocol was repeated for the topological laser on a $N_x=128$ times $N_y=12$ stripe with periodic boundary conditions and $J=5\gamma$. The results} are reported in red squares and triangles in Fig.~\ref{fig:tau_W}. \iac{Simulations were also performed with fully open boundary conditions and gain distributed along the whole edge, yielding the same conclusions. The system size with $N_x=52$ and $N_y=14$ was chosen to have an edge with the same length of 128 sites. The results are shown in the figure by the red $\times$ markers.}

\iac{In contrast to the non-topological case}, the behavior of the topological laser \iac{remains quite regular when disorder is introduced and different realizations show very similar features. Temporal and spatial coherence is very well preserved (as apparent in Fig.~S8 \citep{suppl}) until a marked threshold at a value of disorder $W_{thr}\sim J$ on the order of the topological gap of the underlying Harper--Hofstadter model. Beyond this value, spatial and temporal coherence are rapidly lost. } 

\iac{More precisely, for small $W$, disorder has a minor impact at all  and the laser field in the chiral edge mode is able to continuously travel around the system\iaciac{. For intermediate values of the disorder strength, but still below the threshold,} there is a surprising and systematic enhancement of the temporal coherence.} The coherence time \iac{remains} anyway \iac{well below} than the ultimate Bogoliubov--Schawlow--Townes bound (\ref{eq:BogoDrift}) which for \iac{our parameters and $W=0$} is $\tau_{ST}/\tau_c \simeq 2.7$.
\iac{A tentative interpretation of these observations} is that disorder \iac{partially} hampers the \iac{KPZ nonlinear} dynamics \iaciac{that is responsible for} the \iac{deviation from the linear Bogoliubov-Schawlow-Townes curve in Fig.~\ref{fig:Nscaling}.}

\iaciac{Finally, it is worth noting that in the previous paragraph we initialized the non-topological laser operation in either the $k^{las}=0$ or the $k^{las}=\frac{\pi}{5}$ modes, finding quantitative different behaviours. In the topological case, instead, the vacuum of the field can be taken as the initial condition with no noticeable change}, demonstrating that disorder \iac{does} not affect the capability of the topological laser to autonomously reach its coherent steady state.


\section{Conclusions}
\label{sec:conclu}
In this work we have investigated the spatio-temporal coherence properties of  arrays of coupled laser resonators, focusing on analogies and differences between a non-topological \iac{one-dimensional} chain and chiral edge-state lasing in a 2D topological Harper--Hofstadter lattice.

In the \iac{non-topological} case we have highlighted the crossover for growing observation times from a Kardar-Parisi-Zhang scaling to a \iac{Schawlow-Townes-like} phase-diffusion regime. Also, \iac{for a growing system size} the \iac{long-time} phase diffusion rate displays a crossover from \iac{the standard} single-mode Schawlow-Townes linewidth to a faster decoherence determined by the nonlinear \iac{and multi-mode} dynamics of spatial fluctuations.
Provided one reasons in the reference frame moving at the group velocity of the chiral mode, these results are found to directly apply to the chiral laser emission in the edge states of a topological laser device. 
\ivan{In particular, in the Schawlow-Townes regime the topological laser emission is characterized by a transverse Petermann factor very close to one: this  further clarifies the nature of the topological localization on the edge, proving that the coherence is  not affected by the geometry of the cavity, as it \iaciac{instead} occurs for lasing in open resonators. For all practical purposes the coherence time is then  determined by the total number of photons in the device, which is the optimal scenario.}

While for clean samples the spatio-temporal correlations behave very similarly for the topological and trivial devices, topological protection entails a much larger resilience to inhomogeneities.
For the \iac{non-topological one-dimensional} chain, static disorder is \iac{in fact} able to spatially localize the lasing mode and/or break it into several disconnected and incoherent pieces. \iac{On} the other hand, the chiral motion of the edge state of a topological laser device is able to maintain the spatial and temporal coherence across the whole sample up to \iac{much larger} values of the disorder strength of the order of the topological gap.
These results open exciting perspectives both for technological applications and for studies of fundamental physics \iac{using topological lasers}. 

From the theoretical point of view, ongoing research includes the classification of the different Kardar-Parisi-Zhang universality subclasses for our model \citep{squizzato2018}, the extension of our study to class-B semiconductor lasers by explicitly including the carrier dynamics \citep{longhi2018} and \iac{different kinds of optical nonlinearities that may lead to further decoherence processes. On a longer run, a natural step involves the generalization of the} topological laser concept to photonic lattices with different dimensionalities and different band topologies~\citep{ozawa2019}.

 From the experimental side, the effectively periodic boundary conditions \iac{naturally enjoyed by a chiral edge mode} are extremely promising to suppress undesired spatial inhomogeneities and boundary effects in experimental studies~\cite{baboux2019} of the critical properties of different non-equilibrium statistical models. 

 On the application side, we \iac{demonstrated a level of coherence} quantitatively comparable or even better than the corresponding \iac{non-topological device, with the crucial advantage that the coherence of a topological laser is robust against the presence of static disorder. While a related robustness result was established in~\cite{harari2018} for the slope efficiency of the laser device, our crucial result is that it also applies to the coherence properties. This confirms the strong promise that topological laser hold for practical opto-electronic applications.}

\section*{Acknowledgements}
We are grateful to Jacqueline Bloch, Leonie Canet,  Alessio Chiocchetta, Aurelian Loirette--Pelous, Matteo Secl{\`i} and Davide Squizzato for useful discussions.
We acknowledge financial support from the European Union FET-Open grant ``MIR-BOSE'' (n. 737017), from the H2020-FETFLAG-2018-2020 project "PhoQuS" (n.820392), and from the Provincia Autonoma di Trento. All numerical calculations were performed using the Julia Programming Language \cite{julia}.

\bibliography{bibliography}

\end{document}


\title{SI: Theory of the coherence of topological lasers}

\author{Ivan Amelio}
\affiliation{INO-CNR BEC Center and Dipartimento di Fisica, Universit{\`a} di Trento, 38123 Povo, Italy}
\author{Iacopo Carusotto}
\affiliation{INO-CNR BEC Center and Dipartimento di Fisica, Universit{\`a} di Trento, 38123 Povo, Italy}


\date{\today}

\maketitle

\section{Derivation of the equation for the phase}

Here we sketch the derivation of the KPZ equation for the phase from the complex Ginzburg--Landau equation. In particular, as usual in the treatment of quasi-condensates, one considers small and fast density fluctuations, but no assumptions on the phase field.
 
Starting with
\begin{equation}
i\partial_t \psi(x) = \left[ (-J_{eff} + i \eta)\nabla^2  + \frac{i}{2} \left( \frac{P_{eff}}{1+n_S |\psi|^2} - \gamma \right) \right] \psi + \sqrt{2D} \xi
\end{equation}
the modulus-phase equations  for $\psi = \sqrt{n_0 + \delta n}e^{i\phi}$ read:
\begin{equation}
-\partial_t \phi =  -\eta \nabla^2 \phi +J_{eff} (\nabla \phi)^2 - \frac{J_{eff}}{2n_0} \nabla^2 \delta n - \frac{\eta}{n_0} \nabla \phi \cdot \nabla \delta n  + \sqrt{\frac{D}{n_0}}\xi_1
\end{equation}.
\begin{equation}
\partial_t \delta n =  \eta \nabla^2 \delta n  - 2 \eta n_0 (\nabla \phi)^2 - 2 J_{eff} n_0 \nabla^2 \phi  -2 J_{eff} \nabla \phi \cdot \nabla \delta n - \Gamma_{eff} \delta n + 2\sqrt{{D n_0}} \xi_2,
\end{equation}
where $\Gamma_{eff} = \gamma(1-\gamma/P_{eff})$. 
The adiabatic and long wavelength approximations lead to take 
\begin{equation}
 \delta n =  \Gamma_{eff}^{-1} \left[  - 2 \eta n_0 (\nabla \phi)^2 - 2 J_{eff} n_0 \nabla^2 \phi \right].
\end{equation}
Substituting \iaciac{this expression} into the phase equation yields a KPZ equation plus linear and nonlinear  terms with four derivatives.

\section{Extraction of the coherence time}

In this paragraph we corroborate the solidity of the results concerning $\tau_c$ and in particular its deviation from the Bogoliubov prediction.

The numerical effort required for the calculation of $\tau_c(N_x)$ grows up rapidly with $N_x$, since one has to access the dynamics at very long times, larger than the KPZ saturation time scaling as $\sim N_x^{3/2}$. For this reason, an intensive analysis of the accuracy of the data points reported in Fig. 2 of the main text  has been restricted to the $N_x=128, J=5\gamma$ case.

\begin{figure}[H]
\centering
\includegraphics[width=1\textwidth]{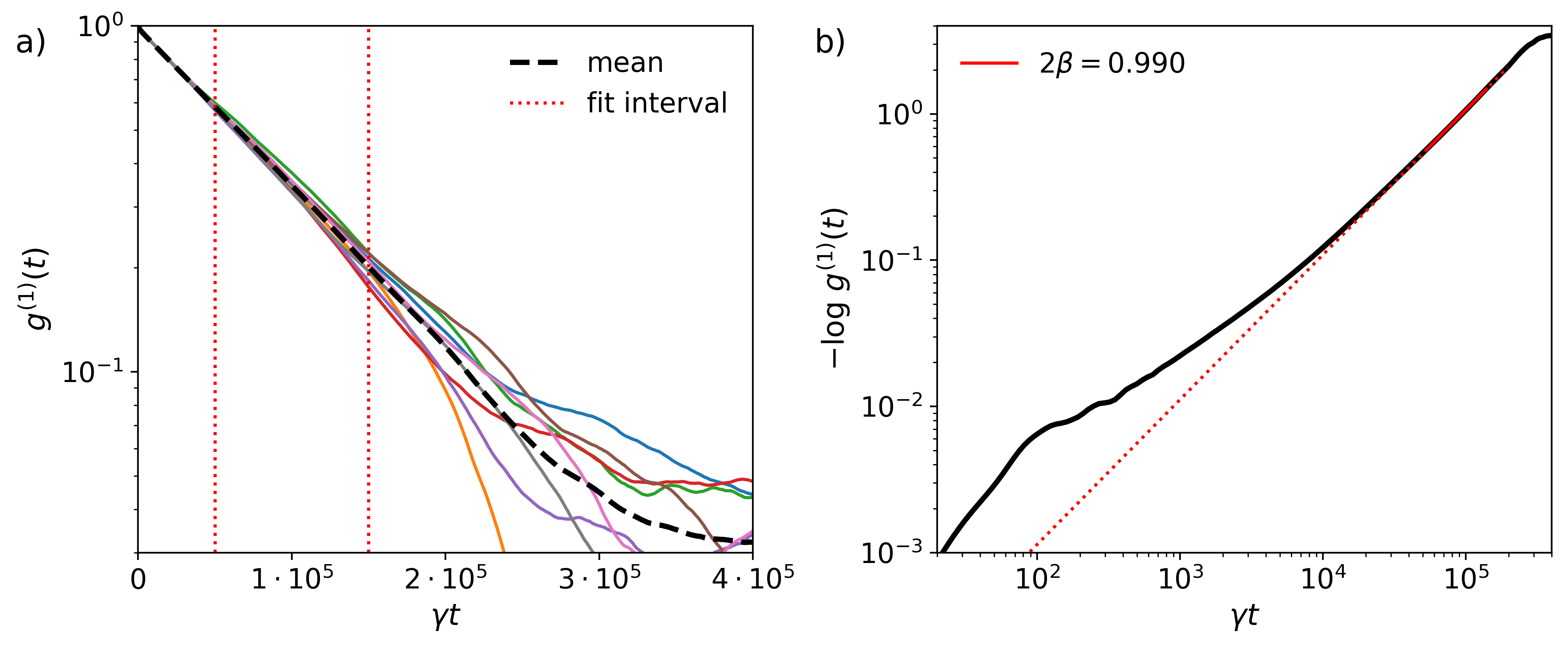}
\caption{{\it Extraction of the coherence time.}  (a) Correlation functions $g^{(1)}(t)$ extracted from 8 runs and their average, for $N_x=128, J=5\gamma$. Linear fitting between red dots yields $\tau_c$. (b) Loglog plot of (minus) the logarithm of averaged $g^{(1)}(t)$. Red line is linear fit of $\log[-\log g^{(1)}(t)]$ with $2\beta \log|t| + B'$, highlighting the Schawlow--Townes regime.
}
\label{fig:8_runs}
\end{figure}

For this dataset we performed 8 runs of duration $T=10^8\gamma^{-1}$, the corresponding correlation functions $g^{(1)}(t)$ being reported in Fig.~\ref{fig:8_runs}.a. Because of finite statistics, the resolution of $g^{(1)}(t)$ is limited and the curves reach a noisy plateaux  at very large times. 
The average of the 8 runs is taken (more specifically the complex functions $\langle \psi^*(x, t) \psi(x, 0) \rangle$ are averaged over all the runs and the absolute modulus is taken afterwards)  and 
we extract $\tau_c$ by a linear fit of $\log g^{(1)}(t)$ in the interval indicated by the red dots, which is of the order of $\tau_c$ itself.
The same is done also for the single runs, resulting in a mean coherence time of $\gamma \tau_c \simeq 9.72 \cdot 10^4$ with a standard deviation of $0.73 \cdot 10^4$, which is approximatively the size of the marker in Fig. 2 of the main text. From the averaged correlation function instead we extracted $\gamma \tau_c \simeq 9.40 \cdot 10^4$. Recall that our goal here is mainly to establish deviations from  the Bogoliubov--Schawlow--Townes prediction, which yields the much larger $\gamma \tau_{ST} \simeq 2.56 \cdot 10^5$. 

The logarithm of the averaged $g^{(1)}(t)$ is plotted in loglog scale in Fig.~\ref{fig:8_runs}.b.
Here we fit the curve in the same interval of Fig.~\ref{fig:8_runs}.a, to prove that $\tau_c$ has been computed by considering large enough times to be in the Schawlow-Townes regime. 

Finally, we argue that this analysis allows to have a rough estimate of the uncertainty of  $\tau_c$ also for the other sizes and couplings, for which we have performed a single run, instead of 8. In particular, for $N_x<128$, the standard deviation computed here is an upper bound for the statistical error done by extracting $\tau_c(N_x)$ from a single simulation of duration $T$.

\section{Temporal coherence of the bulk}

The $U(1)$ global symmetry of the model entails that no force will restore the initial configuration if a global phase shift is kicked to the field. The long time diffusion of the phase is rooted precisely in this Goldstone mode, so that the long time behaviour of the equal-space coherence $g^{(1)}(t)$ does not depend on the lattice site over which it is computed.  Nevertheless, at short times the coherence of the bulk, where the intensity of the field is very small and noise has a major effect, will be very poor.

\begin{figure}[H]
\centering
\includegraphics[width=0.5\textwidth]{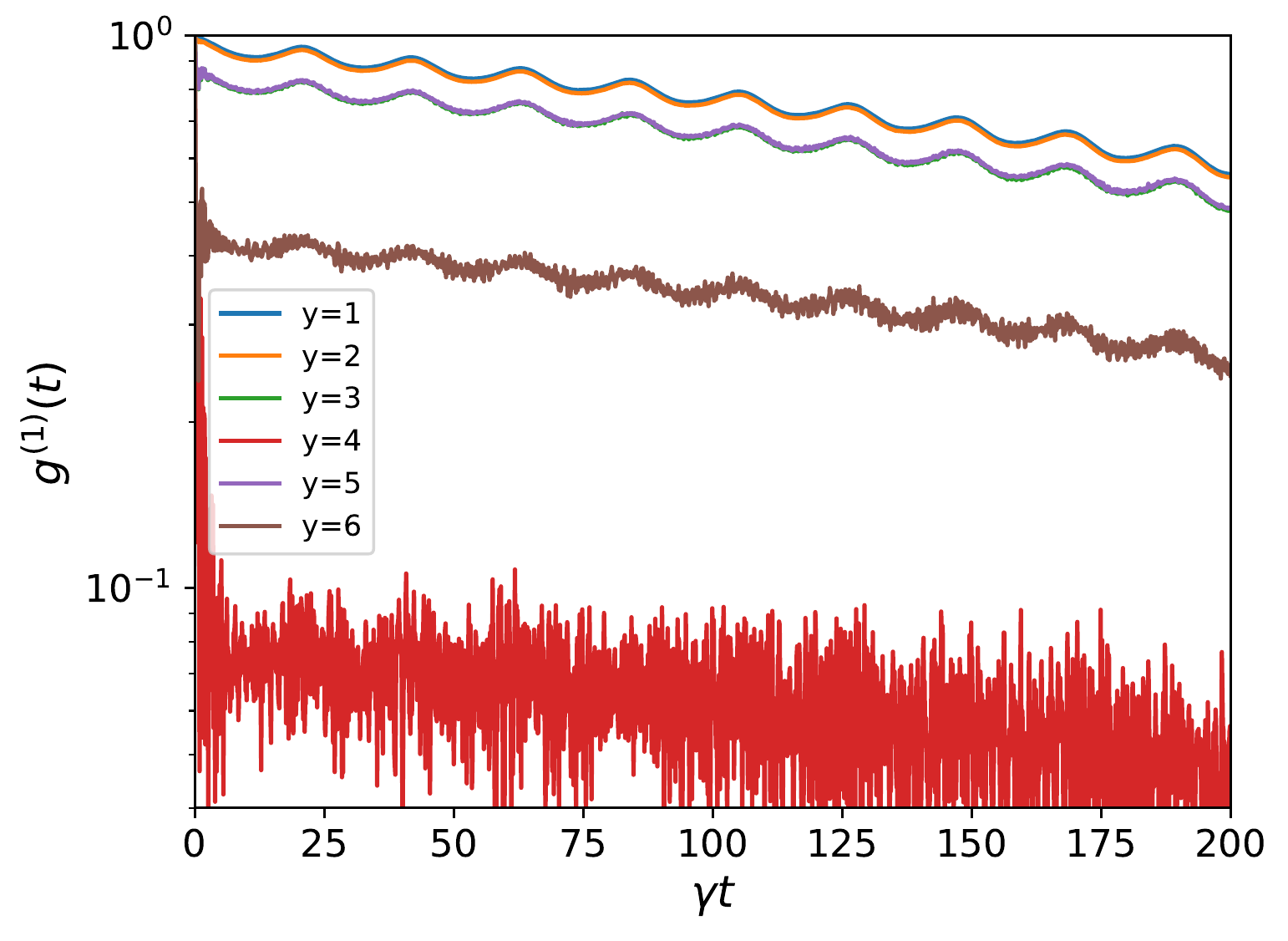}
\caption{{\it Temporal coherence of the bulk.}  
The equal--space time correlation function $g^{(1)}(\vec{x},t)=|\langle \psi^*(\vec{x},t) \psi(\vec{x},0) \rangle|$ is reported in logy scale, for different lattice points $\vec{x}$ (the dependence on the $x$ coordinate is trivial).  As parameters we considered $N_x=128, N_y=12, J=0.5 \gamma, P=2\gamma, n_S=1000, D_{x,y}=\frac{\gamma}{2}(1+\delta_{y,1})$.
}
\label{fig:coherence_bulk}
\end{figure}

How these two simple arguments match it is illustrated in Fig.~\ref{fig:coherence_bulk}, for the same simulation as Fig.'s~(3--5) of the main text. After a sudden collapse at short times, the phase drift $-\log g^{(1)}(t)$ for the points in the bulk grows in time with the same rate as on the edge.
Also, notice that the  modulation given by the chiral trips around the edge  is not important for the coherence on large timescales\iaciac{. For the parameters of this plot, such a modulation} is small because the wavefunction has high spatial coherence along the $x$ direction.

\section{Correlations around a Rectangular Lattice}

In numerical simulations, having the cylindrical configuration is convenient for collecting more statistics, by using translational invariance and recording all sites. In an experiment one can run very long measurements and it may be cumbersome to  record correlations between any pair of sites.
While in the main text we have restricted to periodic boundary conditions, also to have a more pictorial theoretical insight of the phase fluctuations, here we show that no difficulty arises when dealing with open boundary conditions.

\begin{figure}[H]
\centering
\includegraphics[width=1\textwidth]{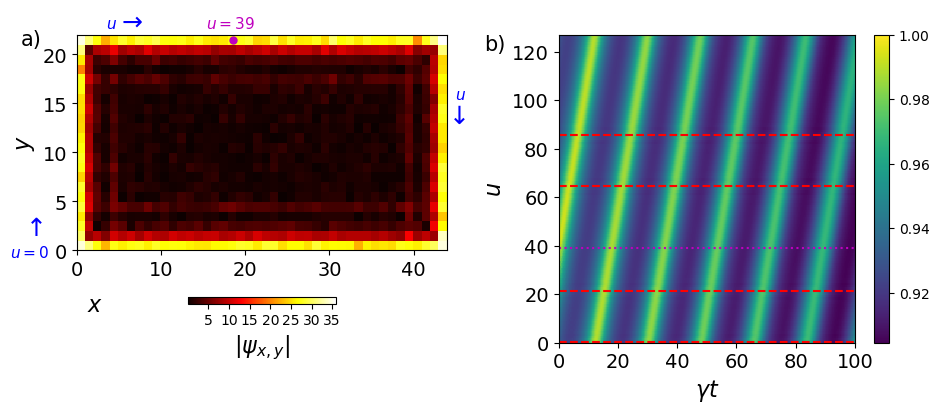} 
\caption{{\it Correlations around a rectangular lattice.}  
(a) Snapshot of the absolute value of the field in a rectangular $N_x=44 \times N_y=22$ lattice with open boundary conditions and amplifying medium along all the edge.  (b) Correlation function $g^{(1)}(u,u_0,t)=\langle \psi^*(u,t) \psi(u_0,t) \rangle$ where $u$ is a coordinate along the edge, with $u_0$ indicated in  magenta. Red dashes correspond to the position of the corners. Apart from the boundary conditions and the lasing point, which has not fixed via the initial conditions but randomly selected by the dynamics, the parameters are the same as Fig.'s~(3--5) of the main text.
}
\label{fig:OBC}
\end{figure}

For a rectangular lattice with the gain on the edge, we have lasing as already reported in the literature \cite{secli2019}, see Fig.~S\ref{fig:OBC}.a. Notice that the density is slighly higher at the corners. By defining an edge coordinate $u$ (which starts from the lower left corner and runs clockwise) and measuring correlation functions with respect to a given point (indicated in magenta), we get the pattern of panel (b), analogous to Fig.~4.c of the main text. 

\section{Passage to comoving frame}

\begin{figure}[H]
\centering
\includegraphics[width=1\textwidth]{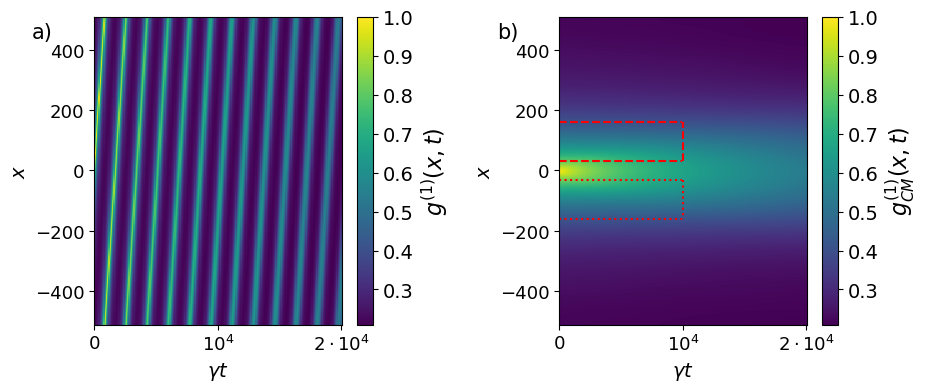}
\caption{{\it Passage to comoving frame.}  
Correlation functions in the lab (a) and comoving frame (b). The data in Fig.~6.b of the main text are  the (properly rescaled) horizontal cuts of the areas enclosed in the red rectangles.
}
\label{fig:to_CM_KPZ}
\end{figure}

We illustrate once again the procedure to assess KPZ  physics in a real-life experiment.
One can compute or measure the correlation functions $g^{(1)}(x,t)$ in the lab frame and get data looking like Fig.~S\ref{fig:to_CM_KPZ}.a below. (The edge of the system can be of whatever shape, as suggested in Fig.~S\ref{fig:OBC}.)
At this point one should fit $v_g$ from the chiral pattern and compute $g_{_{CM}}^{(1)}(x, t) = g^{(1)}(x-v_g t, t)$.

The result of this post-processing is plotted in Fig.~\ref{fig:to_CM_KPZ}.b. In particular, the data in Fig.~6.b of the main text are  the (properly rescaled) horizontal cuts of the two areas enclosed in the two rectangles drawn by red dashes and red dots.

\section{Small-bandgap topological lasing}

It is well known that the edge modes of a topological insulator are  localized and protected against backscattering as long as the energy scales associated with static disorder, dissipation and noise are smaller than the bandgap. 
In the main paper, in order to study numerically KPZ physics, it was convenient to use a small $J$   and intermediate values of noise. In particular, we repeatedly considered $J=0.5\gamma$, in which regime the bare Harper--Hofstadter bands would not be well resolved.

Still, as we show here, topological lasing occurs even in this case, with the edge mode maintaining its localization and resilience to backscattering.

\begin{figure}[H]
\centering
\includegraphics[width=1\textwidth]{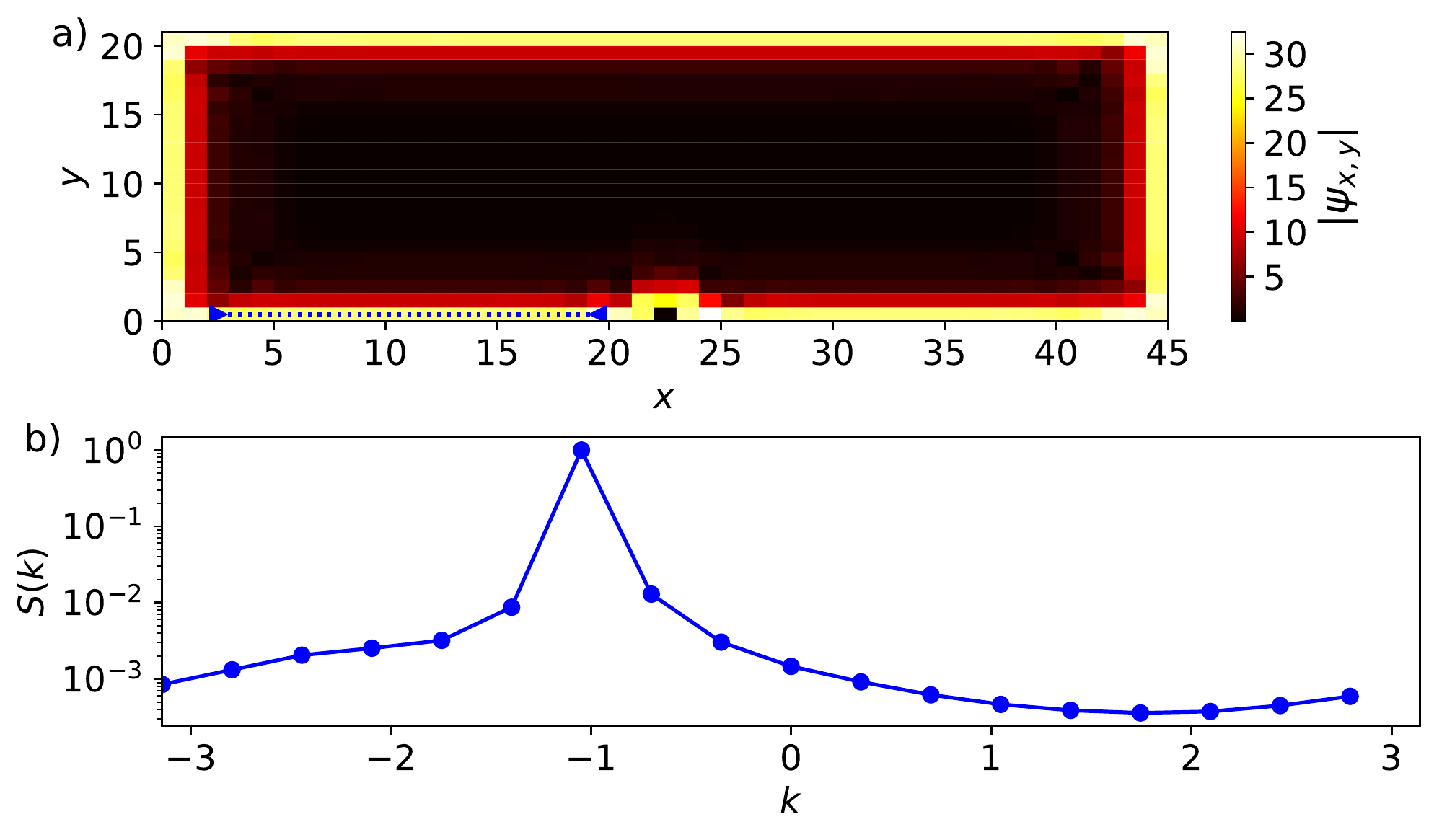}
\caption{{\it Small-bandgap topological lasing.}  
A $45 \times 21$ Harper--Hofstadter lattice with $J=0.5\gamma$ and open boundary conditions is built with a strong defect on an edge. Amplifying the edges result in topological lasing: the modulus of the field is visible in panel (a) \iaciac{and} the momentum-space intensity along the blue dots is plotted in panel (b)\iaciac{Both panels show no trace of} backscattering\iaciac{. In this case} the chiral mode \iaciac{on the lower edge} goes from left to right, since for this negative $k$ the group velocity is positive.}
\label{fig:robust_m10}
\end{figure}

In particular, we consider a strong defect on the boundary of a rectangular lattice (so that also the corners act like defects) and put gain all over the edge of the system. \iaciac{Starting the simulation with a small random seed, the fields reach a steady-state which displays the usual features of a topological laser and is able to go around the strong defect,} as it is shown in Fig.~\ref{fig:robust_m10}. In panel (a) we depict the modulus  of the field, while the absence of backscattering is demonstrated by plotting, in panel (b), the Fourier intensity $S(k)$ of the field along the blue dots of panel (a).  This indicates that the emission is narrow and single-peaked in momentum space, and corresponds to positive group velocity (as \iaciac{it is visible in} Fig.~3.a \iaciac{of} the main text).

\section{Finite size effects in KPZ dynamics}

\begin{figure}[H]
\centering
\includegraphics[width=1\textwidth]{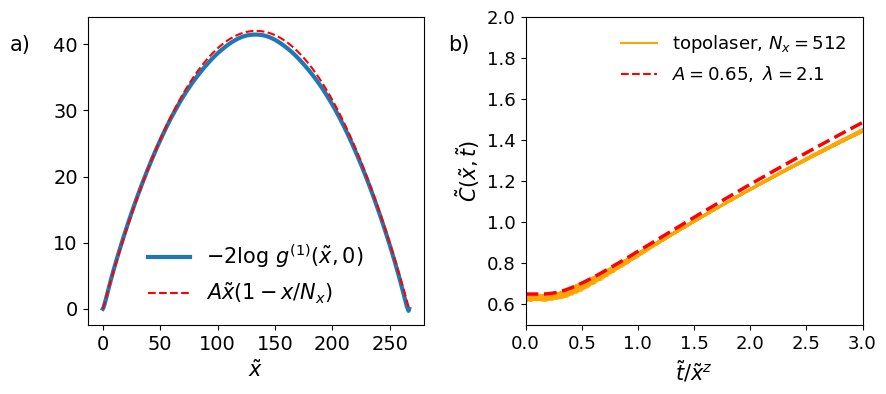}
\caption{{\it Finite--size effects  in KPZ dynamics.}  
We assess finite-size effects by considering the correlation functions of a topological laser of length $N_x=512$ in the KPZ regime (parameters like Fig.~6 of the main text).
(a) Equal-time spatial correlation function. For an infinite system one would have  $-2\log \ g^{(1)}(\tilde{x},0) = A |\tilde{x}|$; finite-size effects are kept into account by adding the Brownian bridge factor.
(b) The correlation functions (orange) collapse, in very good approximation, onto the same universal scaling function of the twice longer system $N=1024$ (red dashes).
}
\label{fig:finite_size}
\end{figure}

In the main text we considered the KPZ dynamics of a $N_x=1024 \times N_y=12$ topological lattice. Here we prove that we can easily take into account  the spatial finite size effects by multiplying the phase-phase correlation functions by a factor $1-x/N_x$. 
This strategy is also adopted in Fig.~6.b of the main text. 
The rational behind this trick is that the probability of having an equal-time jump of $\phi(x)-\phi(0)$ in an infinite system is proportional to the Gaussian $e^{-\frac{\Delta\phi^2}{Ax}}$.  In a finite system  with periodic boundary conditions one has to multiply the probabilities of making the jump by crossing or not the boundary of the system $e^{-\frac{\Delta\phi^2}{Ax}}\times e^{-\frac{\Delta\phi^2}{A(L-x)}}=e^{-\frac{\Delta\phi^2}{Ax(1-x/L)}}$, which is called a Brownian bridge.
With this strategy the theoretical prediction perfectly matches the numerical spatial correlation function, as shown in Fig.~\ref{fig:finite_size}.a.
 
Finally, we checked that the effective KPZ parameters  $\lambda , A$ estimated in the main text \iaciac{do not} depend significantly on the size of the system, in other words the renormalization flow has reached its fixed point.  This is shown in Fig.~\ref{fig:finite_size}.b, where the correlation functions of a \iaciac{shorter} $N_x=512$ long system (orange lines) fall close to the scaling function fitted from the  $N_x=1024$ data (red dashes, the same as Fig.~6.b of the main text).

\section{Disordered lasing phenomenology: examples}

\begin{figure}[H]
\centering
\includegraphics[width=1\textwidth]{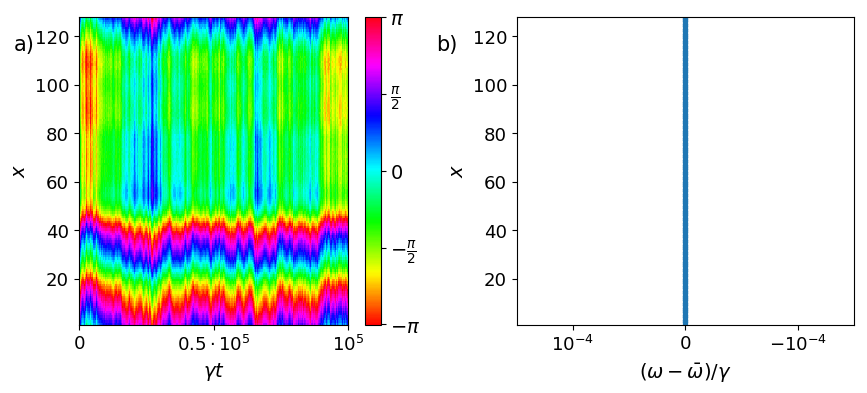}
\includegraphics[width=1\textwidth]{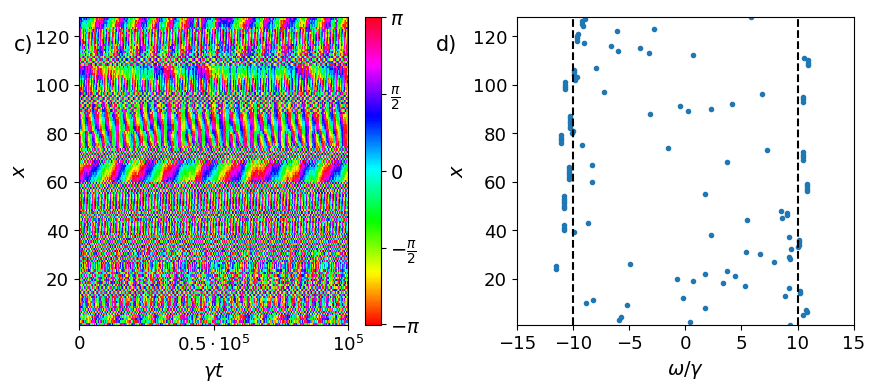}
\caption{{\it \iaciac{Non-topological} disordered 1D laser.}  
For \iaciac{a relatively weak disorder} $W=0.02J$ we report (a) the evolution of the phase of the field corresponding to the blue circle data from Fig.~8 of the main text and 
(b) \iaciac{the} lasing frequency of each site on the edge\iaciac{, measured with respect to the spatial average of the lasing frequencies}. The spatial behavior of the phase is changed by disorder without affecting the spatial coherence, and, very curiously, all sites lase at the same frequency \iaciac{even though} the coherence time is reduced to one tenth.
\iaciac{Panels} (c-d) are the same as \iaciac{panels} (a-b) \iaciac{except for the much stronger disorder} $W=0.32J$: the lattice breaks up in small domains lasing at \iaciac{distinct} frequencies \iaciac{spread in the frequency interval between the} the band edges of the system (vertical black dashes).
}
\label{fig:disorder_trivial}
\end{figure}

\begin{figure}[H]
\centering
\includegraphics[width=1\textwidth]{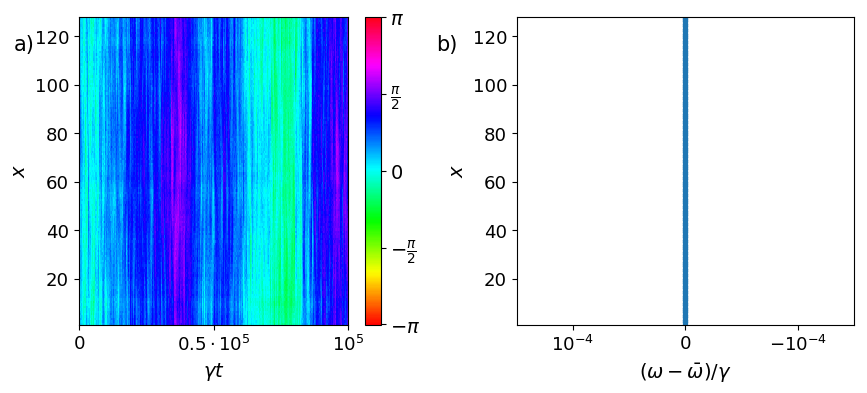}
\includegraphics[width=1\textwidth]{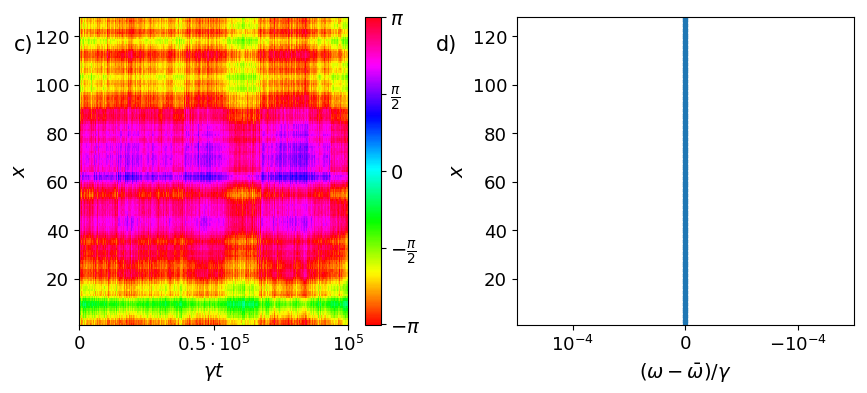}
\caption{{\it Disordered topological laser.}  
For $W=0.02J$ we report (a) the typical evolution of the phase of the field in the slowly varying frame and 
(b) the lasing frequency of each site on the edge (with respect to the spatial average). 
\iaciac{Panels} (c-d) are the same as \iaciac{panels} (a-b) \iaciac{except for the stronger disorder} $W=0.32J$: in spite of some dependence of the phase on $x$ (indicating that the steady-state is not a perfect plane wave $e^{ik_x^{las}x}$) all the lattice sites still lase in a spatially coherent way. Parameters are the same of Fig.~8 of the main text, and data correspond to the red square data series.
}
\label{fig:disorder_topo}
\end{figure}

Even though a detailed study of the lasing dynamics in the presence of static disorder goes well beyond of the scope of this work, we report a small  sample of the \iaciac{different behaviours that can be}  observed in this case \iaciac{for different realizations of disorder}.
In Fig.~\ref{fig:disorder_trivial} and \ref{fig:disorder_topo} we consider respectively the 1D laser (initialized in $k^{las}
_x=0$ and the topological laser, with parameters as in Fig.~8 of the main text.
In the left panels the phase of $\psi_{sl}(x,t)$ is plotted in the slowly varying frame, while in the right panels the frequency of each site $\omega(x)$, defined as the peak of the spectrum of the field on that site, is plotted with respect to the spatial average $\bar{\omega} = \frac{1}{N_x}\sum_x \omega(x)$, with the exception of panel Fig.~\ref{fig:disorder_trivial}.d where \iaciac{the absolute frequency is plotted instead.}.

\bibliography{bibliography}